\documentclass[12pt]{article}
\usepackage{amsfonts}
\usepackage{amssymb}
\usepackage{amsmath}

\begin{document}
\title{The Holevo capacity of infinite dimensional channels and the additivity problem.}
\author{M.E.Shirokov \thanks{Steklov Mathematical Institute, 119991 Moscow,
Russia, e-mail: msh@mi.ras.ru}}
\date{}
\maketitle

\begin{abstract}
The Holevo capacity of arbitrarily constrained infinite
dimensional quantum channel is considered and its properties are
discussed. The notions of input and output optimal average states
are introduced. The continuity properties of the Holevo capacity
with respect to constraint and to channel are explored.

The main result of this paper is the statement that additivity of
the Holevo capacity for all finite dimensional channels implies
its additivity for all infinite dimensional channels with
arbitrary constraints.

\end{abstract}

\section{Introduction}

The Holevo capacity (in what follows, $\chi$-capacity) of a
quantum channel is an important characteristic defining amount of
classical information, which can be transmitted by this channel
using nonentangled encoding and entangled decoding, see e.g.
\cite{H-QI}, \cite{H-c-w-c},\cite{Sch-West}. For additive channels
the $\chi$-capacity coincides with the full classical capacity of
a quantum channel. At present the main interest is focused on
quantum channels between finite dimensional quantum systems. But
having in mind possible applications it is necessary to deal with
infinite dimensional quantum channels, in particular, Gaussian
channels.

In this paper the  $\chi$-capacity for arbitrarily constrained
infinite dimensional quantum channel is considered. It is shown that
despite nonexistence of an optimal ensemble in this case it is
possible to define the notion of the output optimal average state
for such a channel, inheriting important properties of the image of
the average state of an optimal ensemble for finite dimensional
channels (proposition 1).  A "minimax" expression for the
$\chi$-capacity is obtained and an alternative characterization of
the output optimal average state as a minimum point of a lower
semicontinuous function on a compact set is given (proposition 2).

The notion of the $\chi$-function of an infinite dimensional
quantum channel is introduced. It is shown that the
$\chi$-function of an arbitrary channel is a concave lower
semicontinuous function with natural chain properties, having
continuous restriction to any set of continuity of the output
entropy (propositions 3-4). This and the result in \cite{H-Sh-2}
imply continuity of the $\chi$-function for Gaussian channels with
power constraint (example 1). For the $\chi$-function the analog
of Simon's dominated convergence theorem for quantum entropy is
also obtained (corollary 4).

The question of continuity of the $\chi$-capacity as a function of
channel is considered. It is shown that the $\chi$-capacity is
continuous function of channel in the finite dimensional case
while in general it is only lower semicontinuous (theorem 1,
example 2).

The above results make it possible to obtain the infinite
dimensional version of theorem 1 in \cite{H-Sh}, which shows
equivalence of several formulations of the additivity conjecture
(theorem 2).

The main result of this paper is the statement that additivity of
the $\chi$-capacity for all finite dimensional channels implies
its additivity  for all infinite dimensional channels with
arbitrary constraints (theorem 3). This is done in two steps by
using several results (lemma 5, proposition 5 and 6). These
results are also applicable to analysis of individual pairs of
channels as it is demonstrated in the proof of additivity of the
$\chi$-capacity for two arbitrarily constrained infinite
dimensional channels with one of them noiseless or entanglement
breaking (proposition 7).

\section{Basic quantities}

Let $\mathcal{H}$ be a separable Hilbert space,
$\mathfrak{B}(\mathcal{H})$ be the set of all bounded operators on
$\mathcal{H}$ with the cone $\mathfrak{B}_{+}(\mathcal{H})$ of all
positive operators, $\mathfrak{T}( \mathcal{H})$ be the Banach
space of all trace-class operators with the trace norm
$\Vert\cdot\Vert_{1}$ and $\mathfrak{S}(\mathcal{H})$ be the
closed convex subset of $\mathfrak{T}(\mathcal{H})$ consisting of
all density operators on $\mathcal{H}$, which is complete
separable metric space with the metric defined by the trace norm.
Each density operator uniquely defines a normal state on
$\mathfrak{B}(\mathcal{H})$ \cite{B&R}, so, in what follows we
will also for brevity use the term "state". Note that convergence
of a sequence of states to a \textit{state} in the weak operator
topology is equivalent to convergence of this sequence  to this
state in the trace norm \cite{D-A}.

In what follows $\log$ denotes the function on $[0,+\infty)$,
which coincides with the logarithm on $\left( 0,+\infty \right) $
and vanishes at zero. Let $A$ and $B$ be positive trace class
operators. Let $\{|i\rangle\}$ be a complete orthonormal set of
eigenvectors of $A$. The entropy is defined by
$H(A)=-\sum_{i}\langle i|\,A\log A\,|i\rangle$ while the relative
entropy -- as $ H(A\,\|B)=\sum_{i}\langle i|\,(A\log A-A\log
B+B-A)\,|i\rangle$, provided
$\mathrm{ran}A\subseteq\mathrm{ran}B$,\footnote{$\mathrm{ran}$
denotes the closure of the range of an operator in $\mathcal{H}$}
and $H(A\,\Vert B)=+\infty$ otherwise (see \cite{L-3}, \cite{L}
for more detailed definition). The entropy and the relative
entropy are nonnegative lower semicontinuous (in the trace-norm
topology) concave and convex functions of their arguments
correspondingly \cite{L},\cite{O&P},\cite{W}. We will use the
following inequality
\begin{equation}\label{rel-entr-ineq}
H(\rho\Vert\,\sigma)\geq\textstyle\frac{1}{2}\|\rho-\sigma\|_{1}^{2},
\end{equation}
which holds for arbitrary states $\rho$ and $\sigma$ in
$\mathfrak{S}(\mathcal{H})$ \cite{O&P}.

Arbitrary finite collection $\{\rho _{i}\}$ of states in
$\mathfrak{S}(\mathcal{H})$ with corresponding set of
probabilities $\{\pi _{i}\}$ is called \textit{ensemble} and is
denoted by $\Sigma=\{\pi _{i},\rho _{i}\}$. The state
$\bar{\rho}=\sum_{i}\pi _{i}\rho _{i}$ is called \textit{the
average state} of the above ensemble. Following \cite{H-Sh-2} we
treat an arbitrary Borel probability measure $\pi$ on
$\mathfrak{S}(\mathcal{H})$ as \textit{generalized ensemble} and
the \textit{barycenter}  of the measure $\pi$ defined by the
Pettis integral
\[
\bar{\rho}(\pi )=\int\limits_{\mathfrak{S}(\mathcal{H})}\rho \pi
(d\rho )
\]
as the average state of this ensemble. In this notations the
conventional ensembles correspond to measures with finite support.
For arbitrary subset $\mathcal{A}$ of $\mathfrak{S}(\mathcal{H})$
we denote by $\mathcal{P}_{\mathcal{A}}$ the set of all
probability measures with barycenters contained in $\mathcal{A}$.

For arbitrary finite set of ensembles
$\{\{\pi^{k}_{i},\rho^{k}_{i}\}_{i=1}^{n(k)}\}_{k=1}^{m}$ and
arbitrary probability distribution $\{\lambda_{k}\}_{k=1}^{m}$
consider the ensemble consisting of $\sum_{k=1}^{m}n(k)$ states
$\{\rho^{k}_{i}\}_{ki}$ with the corresponding probabilities
$\{\lambda_{k}\pi^{k}_{i}\}_{ki}$. We will call this ensemble convex
combination of the above ensembles and denote it by
$\sum_{i=1}^{m}\lambda_{k}\{\pi^{k}_{i},\rho^{k}_{i}\}_{i=1}^{n(k)}$.
By using the relation between the notion of an ensemble and the
notion of a probability measure one can say that a convex
combination of ensembles corresponds to a convex combination of
measures corresponding to these ensembles.

In analysis of the $\chi$-capacity we shall use Donald's identity
\cite{Don},\cite{O&P}
\begin{equation}\label{Donald's identity}
\sum_{i=1}^{n}\pi_{i}H(\rho_{i}\Vert
\hat{\rho})=\sum_{i=1}^{n}\pi_{i}H(\rho_{i}\Vert
\bar{\rho})+H(\bar{\rho}\Vert \hat{\rho}),
\end{equation}
which holds for arbitrary ensemble $\{\pi _{i},\rho _{i}\}$ of $n$
states with the average state $\bar{\rho}$ and arbitrary state
$\hat{\rho}$.

Let $\mathcal{H},\mathcal{H}^{\prime }$ be a pair of separable
Hilbert spaces which we shall call correspondingly input and
output space. A channel $\Phi $ is a linear positive trace
preserving map from $\mathfrak{T}(\mathcal{ H })$ to
$\mathfrak{T}(\mathcal{H}^{\prime })$ such that the dual map $\Phi
^{\ast }:\mathfrak{B}(\mathcal{H}^{\prime
})\mapsto\mathfrak{B}(\mathcal{H})$ (which exists since $\Phi $ is
bounded \cite{Dev-2}) is completely positive. Let $\mathcal{A}$ be
an arbitrary closed subset of $\mathfrak{S}(\mathcal{H})$. We
consider constraint on input ensemble $\{\pi _{i},\rho _{i}\}$,
defined by the requirement $\bar{\rho}\in \mathcal{A}$. The
channel $\Phi $ with this constraint is called the $\mathcal{A}$
-\textit{constrained} channel. We define the
$\chi$-\textit{capacity} of the $\mathcal{A}$-constrained channel
$\Phi$ as (cf.\cite{H},\cite{H-c-w-c},\cite{H-Sh})
\begin{equation}\label{ccap-1}
\bar{C}(\Phi ;\mathcal{A})=\sup_{\bar{\rho}\in
\mathcal{A}}\chi_{\Phi}(\{\pi_{i},\rho_{i}\}),
\end{equation}
where
$$
\chi_{\Phi}(\{\pi_{i},\rho_{i}\})=\sum_{i}\pi_{i}H(\Phi (\rho
_{i})\|\Phi(\bar{\rho})).
$$

In \cite{H-Sh-2} it is shown that the $\chi$-capacity of the
$\mathcal{A}$-constrained channel $\Phi$ can be also defined by
\begin{equation}\label{ccap-2}
\bar{C}(\Phi;\mathcal{A})=\sup_{\pi\in\mathcal{P}_{\mathcal{A}}}\int\limits_{\mathfrak{S}(
\mathcal{H})}H(\Phi(\rho)\Vert\Phi(\bar{\rho}(\pi)))\pi(d\rho),
\end{equation}
which means coincidence of the above supremum over all measures in
$\mathcal{P}_{\mathcal{A}}$ with the supremum over all measures in
$\mathcal{P}_{\mathcal{A}}$ with finite support.

The $\chi$-capacity $\bar{C}(\Phi;\mathfrak{S}(\mathcal{H}))$ of
the unconstrained channel $\Phi$ is also denoted by
$\bar{C}(\Phi)$.

The $\chi$-function of the channel $\Phi$ is defined by
\begin{equation}\label{chi-fun-def}
\chi_{\Phi}(\rho)=\bar{C}(\Phi
;\{\rho\})=\sup_{\sum_{i}\pi _{i}\rho _{i}=\rho}\sum_{i}\pi_{i}H(\Phi (\rho
_{i})\|\Phi(\rho)).
\end{equation}
The $\chi$-function of finite dimensional channel $\Phi$ is a
continuous concave function on $\mathfrak{S}(\mathcal{H})$
\cite{H-Sh}. The properties of the $\chi$-function of arbitrary
infinite dimensional channel $\Phi$ are considered in section 4.

\section{The optimal average state}

It is known fact that for an arbitrary finite dimensional channel
$\Phi$ and arbitrary closed set $\mathcal{A}$ there exists an
optimal ensemble $\{\pi _{i},\rho _{i}\}$ on which the supremum in
the definition (\ref{ccap-1}) of the $\chi$-capacity is achieved
\cite{Dev-1},\cite{Sch-West-1}. The image of the average state of
this optimal ensemble plays an important role in the analysis of
finite dimensional channels \cite{H-Sh}.

For general infinite dimensional constrained channel there are no
reasons for existence of an optimal ensemble (with finite number of
states). In this case it is natural to introduce the notion of
optimal generalized ensemble (optimal measure) on which the supremum
in the definition (\ref{ccap-2}) of the $\chi$-capacity is achieved.
In \cite{H-Sh-2} the sufficient condition for existence of an
optimal measure for infinite dimensional constrained channel is
obtained and the example of the channel with no optimal measure is
given.

The aim of this section is to show that even in the case of
nonexistence of optimal generalized ensemble we can define the
notion of output optimal average state, inheriting the basic
properties of the image of the average state of an optimal
ensemble for a finite dimensional constrained channel. Using this
notion we can generalize some results of \cite{H-Sh} to the
infinite dimensional case.

Note first that Donald's identity (\ref{Donald's identity})
implies the following observation.

\textbf{Lemma 1.} \textit{Let
$\{\{\pi^{k}_{i=1},\rho^{k}_{i}\}_{i}^{n(k)}\}_{k=1}^{m}$ be a
finite set of ensembles and $\{\lambda_{k}\}_{k=1}^{m}$ be a
probability distribution. Then
$$
\chi_{\Phi}\left(\sum_{k=1}^{m}\lambda_{k}\{\pi^{k}_{i},\rho^{k}_{i}\}_{i=1}^{n(k)}\right)=
\sum_{k=1}^{m}\lambda_{k}\chi_{\Phi}\left(\{\pi^{k}_{i},\rho^{k}_{i}\}_{i=1}^{n(k)}\right)
+\chi_{\Phi}\left(\{\lambda_{k},\bar{\rho}_{k}\}_{k=1}^{m}\right),
$$
where $\bar{\rho}=\sum_{i=1}^{m}\lambda_{k}\bar{\rho}_{k}$ is the
average state of the ensemble
$\sum_{k=1}^{m}\lambda_{k}\{\pi^{k}_{i},\rho^{k}_{i}\}_{i=1}^{n(k)}$.}

\textit{In the case $m=2$ for arbitrary $\lambda\in [0;1]$ the
following inequality holds}
$$
\begin{array}{c}
\chi_{\Phi}\left(\lambda\{\pi^{1}_{i},\rho^{1}_{i}\}_{i=1}^{n(1)}+
(1-\lambda)\{\pi^{2}_{i},\rho^{2}_{i}\}_{i=1}^{n(2)}\right)\\\\\geq
\lambda\chi_{\Phi}\left(\{\pi^{1}_{i},\rho^{1}_{i}\}_{i=1}^{n(1)}\right)+
(1-\lambda)\chi_{\Phi}\left(\{\pi^{2}_{i},\rho^{2}_{i}\}_{i=1}^{n(2)}\right)
+\frac{\lambda(1-\lambda)}{2}\|\Phi(\bar{\rho}_{2})-\Phi(\bar{\rho}_{1})\|_{1}^{2}.
\end{array}
$$

\textit{Proof.} By definition
$$
\chi_{\Phi}\left(\sum_{k=1}^{m}\lambda_{k}\{\pi^{k}_{i},\rho^{k}_{i}\}_{i=1}^{n(k)}\right)=
\sum_{k=1}^{m}\lambda_{k}\sum_{i=1}^{n(k)}\pi^{k}_{i}
H\left(\Phi(\rho^{k}_{i})\|\Phi(\bar{\rho})\right).
$$
Applying Donald's identity (\ref{Donald's identity}) to the each
inner sum in the right side of the above expression we obtain the
main identity of the lemma.

To prove the inequality in the case $m=2$ it is sufficient to
apply inequality (\ref{rel-entr-ineq}) for the below estimation of
the relative entropies in the main identity of the lemma:
$$
\begin{array}{c}
\lambda
H\left(\Phi(\bar{\rho}_{1})\|\Phi\left(\lambda\bar{\rho}_{1}+(1-\lambda)\bar{\rho}_{2}\right)\right)+
(1-\lambda)
H\left(\Phi(\bar{\rho}_{2})\|\Phi\left(\lambda\bar{\rho}_{1}+(1-\lambda)\bar{\rho}_{2}\right)\right)\\\\\geq
\frac{1}{2}\lambda\|(1-\lambda)\Phi(\bar{\rho}_{2}-\bar{\rho}_{1})\|_{1}^{2}+
\frac{1}{2}(1-\lambda)\|\lambda\Phi(\bar{\rho}_{2}-\bar{\rho}_{1})\|_{1}^{2}\\\\=
\frac{1}{2}\lambda(1-\lambda)\|\Phi(\bar{\rho}_{2})-\Phi(\bar{\rho}_{1})\|_{1}^{2}.
\square
\end{array}
$$

Despite possible nonexistence of optimal ensemble for the
$\mathcal{A}$-constrained channel $\Phi$ the definition of the
$\chi$-capacity implies existence of a sequence of ensembles with
the following properties.

\textbf{Definition 1.} \textit{A sequence of ensembles $\{\pi
_{i}^{k},\rho _{i}^{k}\}$ with the average $\,\bar{\rho}^{k}\in
\mathcal{A}$ such that
$$
\lim_{k\rightarrow+\infty}\chi _{\Phi }(\{\pi _{i}^{k},\rho
_{i}^{k}\})=\bar{C}(\Phi
;\mathcal{A})
$$
is called approximating sequence for the $\mathcal{A}$-constrained
channel $\Phi$.}

\textit{A state $\bar{\rho}$ is called input optimal average state
for the $\mathcal{A}$-constrained channel $\Phi$ if this state
$\bar{\rho}$ is a limit of the sequence of the average states of
some approximating sequence of ensembles for the
$\mathcal{A}$-constrained channel $\Phi$.}

This definition admits that input optimal average state may not
exist or may not be unique. If there exists an optimal measure for
the $\mathcal{A}$-constrained channel $\Phi$ then its barycenter
is an input optimal average state. It follows from lemma 1 and
proposition 1 in \cite{H-Sh-2}. Existence of an input optimal
average state is also a sufficient condition for existence of an
optimal measure for the $\mathcal{A}$-constrained channel $\Phi$
if the restriction of the output entropy to the set $\mathcal{A}$
is continuous at this state \cite{H-Sh-2}.

Despite possible nonexistence of partial limits of the sequence of
the average states of a particular approximating sequence the
following proposition guarantees convergence of the sequence of
their images.

\textbf{Proposition 1.} \textit{Let $\mathcal{A}$ be convex subset
of $\mathfrak{S}(\mathcal{H})$ such that
$\bar{C}(\Phi;\mathcal{A})<+\infty$. Then there exists the unique
state $\,\Omega(\Phi,\mathcal{A})$ in
$\,\mathfrak{S}(\mathcal{H}')$ such that
\[
\sup_{\sum_{j}\mu _{j}\sigma _{j}\in\mathcal{A}}\sum_{j}\mu
_{j}H(\Phi (\sigma _{j})\Vert
\Omega(\Phi,\mathcal{A}))=\bar{C}(\Phi
;\mathcal{A}),
\]
(the supremum is over all ensembles $\{\mu _{j},\sigma_{j}\}$ with
the average state $\bar{\sigma}\in \mathcal{A}$).}

\textit{For arbitrary approximating sequence of ensembles
$\{\pi_{i}^{k},\rho _{i}^{k}\}$ for the $\mathcal{A}$-constrained
channel $\Phi$ there exists}
$$
\lim_{k\rightarrow+\infty}\Phi(\bar{\rho}_{k})=\Omega(\Phi,\mathcal{A}).
$$

\textit{Proof.} Show first that for arbitrary approximating sequence
of ensembles $\left\{\Sigma_{k}=\{\pi_{i}^{k},\rho
_{i}^{k}\}_{i=1}^{n(k)}\right\}$ for the $\mathcal{A}$-constrained
channel $\Phi$ the sequence $\{\Phi(\bar{\rho}_{k})\}$ converges to
a particular state in $\mathfrak{S}(\mathcal{H}')$. By definition of
an approximating sequence for arbitrary $\varepsilon>0$ there exists
$N_{\varepsilon}$ such that
$\chi_{\Phi}(\Sigma_{k})>\bar{C}(\Phi;\mathcal{A})-\varepsilon$ for
all $k\geq N_{\varepsilon}$. By lemma 1 (with $m=2$ and
$\lambda=1/2$) for all $k_{1}\geq N_{\varepsilon}$ and $k_{2}\geq
N_{\varepsilon}$ we have
$$
\begin{array}{c}
\bar{C}(\Phi;\mathcal{A})-\varepsilon\leq
\frac{1}{2}\chi_{\Phi}(\Sigma_{k_{1}})+\frac{1}{2}\chi_{\Phi}(\Sigma_{k_{2}})\\\\
\!\!\leq\chi_{\Phi}\left(\frac{1}{2}\Sigma_{k_{1}}+\frac{1}{2}\Sigma_{k_{2}}\right)-
\frac{1}{8}\|\Phi(\bar{\rho}_{k_{2}})-\Phi(\bar{\rho}_{k_{1}})\|^{2}\leq
\bar{C}(\Phi;\mathcal{A})-
\frac{1}{8}\|\Phi(\bar{\rho}_{k_{2}})-\Phi(\bar{\rho}_{k_{1}})\|^{2},
\end{array}
$$
and hence
$\|\Phi(\bar{\rho}_{k_{2}})-\Phi(\bar{\rho}_{k_{1}})\|<\sqrt{8\varepsilon}$.
Thus the sequence $\{\Phi(\bar{\rho}_{k})\}$ is a Cauchy sequence
and hence it converges to a particular state $\rho'$ in
$\mathfrak{S}(\mathcal{H}')$.

Let $\{\mu _{j},\sigma _{j}\}_{j=1}^{m}$ be an arbitrary ensemble
with the average $\bar{\sigma}\in \mathcal{A}$. Consider the
family of ensembles
\[
\Sigma^{\eta}_{k}=(1-\eta)\{\pi_{i}^{k},\rho
_{i}^{k}\}_{i=1}^{n(k))}+\eta\{\mu _{j},\sigma
_{j}\}_{j=1}^{m},\quad \eta\in [0,1], k\in \mathbb{N}
\]
with the average states $\bar{\rho}_{k}^{\eta}$. By convexity of
$\mathcal{A}$ we have $\bar{\rho}_{k}^{\eta}\in\mathcal{A}$ for all
$\eta\in [0,1]$ and $k\in \mathbb{N}$. By the above observation
\begin{equation}\label{limit-exp}
\lim_{k\rightarrow+\infty}\Phi(\bar{\rho}_{k}^{\eta})=(1-\eta)\rho'+\eta\Phi(\bar{\sigma}).
\end{equation}

By definition
\begin{equation}
\chi _\Phi\left(\Sigma_{k}^{\eta} \right)=(1-\eta
)\sum_{i=1}^{n(k)}\pi^{k}_{i}H(\Phi (\rho^{k}_{i})\Vert \Phi
(\bar{\rho}_{k}^{\eta} ))+\eta \sum_{j=1}^m\mu _jH(\Phi
(\sigma_{j})\Vert \Phi (\bar{\rho}_{k}^{\eta})). \label{m-chi}
\end{equation}
Since $\bar{C}(\Phi;\mathcal{A})<+\infty$ the both sums in the right
side of this expression are finite. Applying Donald's identity
(\ref{Donald's identity}) to the first sum  we obtain
\[
\sum_{i=1}^{n(k)}\pi^{k}_{i}H(\Phi (\rho^{k}_{i})\Vert \Phi
(\bar{\rho}_{k}^{\eta}))=\chi _\Phi (\Sigma_{k}^{0})+H(\Phi
(\bar{\rho}_{k})\Vert \Phi(\bar{\rho}_{k}^{\eta})).
\]
Substitution of the above expression into (\ref{m-chi}) gives
$$
\begin{array}{c}
\chi _\Phi \left( \Sigma_{k}^{\eta} \right) =\chi _\Phi
(\Sigma_{k}^{0})+(1-\eta )H(\Phi (\bar{\rho}_{k})\Vert \Phi
(\bar{\rho}_{k}^{\eta}))
\\
\\
+\eta \left[ \sum\limits_{j=1}^m\mu _jH(\Phi (\sigma _j)\Vert \Phi
(\bar{\rho}_{k}^{\eta}))-\chi _\Phi (\Sigma_{k}^{0})\right].
\end{array}
$$
Due to nonnegativity of the relative entropy it follows that
\begin{equation} \sum\limits_{j=1}^m\mu _jH(\Phi(\sigma
_j)\Vert \Phi (\bar{\rho}^{k}_{\eta}
))\leq\eta^{-1}\left[\chi_{\Phi}\left(
\Sigma_{k}^{\eta}\right)-\chi_{\Phi}\left( \Sigma^{k}
_{0}\right)\right]+\chi_{\Phi}\left(\Sigma_{k}^{0}\right),\;\;
\eta\neq 0. \label{chi-ineq-1}
\end{equation}
By definition of the approximating sequence we have
\begin{equation}\label{a-p-exp}
\lim_{k\rightarrow+\infty}\chi_{\Phi}\left(
\Sigma_{k}^{0}\right)=\bar{C}(\Phi
;\mathcal{A})\geq \chi_{\Phi}\left(
\Sigma_{k}^{\eta}\right)
\end{equation}
for all $k$. It follows that
\begin{equation}
\liminf_{\eta\rightarrow+0}\,\liminf_{k\rightarrow+\infty}\,\eta^{-1}\left[\chi_{\Phi}\left(
\Sigma_{k}^{\eta}\right)-\chi_{\Phi}\left( \Sigma^{k}
_{0}\right)\right]\leq 0 \label{d-lim-exp}
\end{equation}

By lower semicontinuity of the relative entropy (\ref{limit-exp}),
(\ref{chi-ineq-1}), (\ref{a-p-exp}) and (\ref{d-lim-exp}) imply
$$
\sum\limits_{j=1}^m\mu_{j}H(\Phi(\sigma_{j})\Vert\rho'))\leq
\liminf_{\eta\rightarrow+0}\,\liminf_{k\rightarrow+\infty}
\sum\limits_{j=1}^m\mu_{j}H(\Phi(\sigma_{j})\Vert \Phi
(\bar{\rho}^{k}_{\eta}))\leq \bar{C}(\Phi
;\mathcal{A}).
$$

This proves that
\begin{equation}\label{C-exp}
\sup_{\sum_{j}\mu _{j}\sigma _{j}\in\mathcal{A}}\sum_{j}\mu
_{j}H(\Phi (\sigma _{j})\Vert \rho')\leq\bar{C}(\Phi
;\mathcal{A}),
\end{equation}
To prove the converse inequality consider an approximating
sequence $\{\pi_{i}^{k},\rho_{i}^{k}\}$. Applying Donald's
identity (\ref{Donald's identity}) we obtain
$$
\sum_{i}\pi_{i}^{k}H(\Phi (\rho
_{i}^{k})\Vert\rho')=\sum_{i}\pi_{i}^{k}H(\Phi (\rho
_{i}^{k})\Vert
\Phi(\bar{\rho}^{k}))+H(\Phi(\bar{\rho}^{k})\Vert\rho').
$$
By the approximating property of the sequence  $\{\pi
_{i}^{k},\rho _{i}^{k}\}$ the first term in the right side tends
to $\bar{C}(\Phi;\mathcal{A})$ as $k\rightarrow+\infty$, while the
second is nonnegative. This proves $"\geq"$ and, hence, $"="$ in
(\ref{C-exp}).

By inequality (\ref{rel-entr-ineq}) and the below lemma 2
inequality (\ref{C-exp}) implies that for arbitrary approximating
sequence of ensembles $\{\mu_{j}^{k},\sigma_{j}^{k}\}$ for the
$\mathcal{A}$-constrained channel $\Phi$ the corresponding
sequence $\Phi(\bar{\sigma}_{k})$ converges to the state $\rho'$.
Thus this state $\rho'$ does not depend on the choice of an
approximating sequence, so, it is determined only by the channel
$\Phi$ and by the constraint set $\mathcal{A}$. Denote this state
by $\Omega(\Phi,\mathcal{A})$. Lemma 2 implies also that
$\rho'=\Omega(\Phi,\mathcal{A})$ is the unique state for which
equality in (\ref{C-exp}) holds. $\square$

\textbf{Lemma 2.} \textit{Let $\mathcal{A}$ be a set such that
$\bar{C}(\Phi;\mathcal{A})<+\infty$ and $\rho'$ be a state in
$\mathfrak{S}(\mathcal{H}')$ such that
\[
\sum_{j}\mu _{j}H(\Phi (\sigma _{j})\Vert\,\rho')\leq \bar{C}(\Phi
;\mathcal{A})
\]
for arbitrary ensemble $\{\mu _{j},\sigma _{j}\}$ with the average
$\bar{\sigma}\in \mathcal{A}$. Then for arbitrary approximating
sequence $\{\pi _{i}^{k},\rho _{i}^{k}\}$ of ensembles for the
$\mathcal{A}$-constrained channel $\Phi$ with the corresponding
sequence of average states $\bar{\rho}_{k}$ there exists
$\lim_{k\rightarrow+\infty}H(\Phi(\bar{\rho}_{k})\|\rho')=0$.}

\textit{Proof.} Let $\{\pi _{i}^{k},\rho _{i}^{k}\}$ an
approximating sequence of ensembles with the corresponding
sequence of the average states $\bar{\rho}^{k}$. By assumption we
have
\[
\sum_{i}\pi_{i}^{k}H(\Phi (\rho _{i}^{k})\Vert\,\rho')\leq
\bar{C}(\Phi
;\mathcal{A}).
\]
Applying Donald's identity (\ref{Donald's identity}) to the left
side we obtain
\begin{equation}\label{D-decomp}
\sum_{i}\pi_{i}^{k}H(\Phi (\rho
_{i}^{k})\Vert\,\rho')=\sum_{i}\pi_{i}^{k}H(\Phi (\rho
_{i}^{k})\Vert \Phi
(\bar{\rho}_{k}))+H(\Phi(\bar{\rho}_{k})\Vert\,\rho')
\end{equation}
From the two above expressions we have
$$
H(\Phi (\bar{\rho}^{k})\Vert\,\rho')\leq \bar{C}(\Phi
;\mathcal{A})-\sum_{i}\pi_{i}^{k}H(\Phi (\rho _{i}^{k})\Vert \Phi
(\bar{\rho}^{k}))
$$
But the right side tends to zero as $k$ tends to infinity due to
the approximating property of the sequence
$\{\pi_{i}^{k},\rho_{i}^{k}\}$. $\square$

Proposition 1 shows in particular that if the set of input average
states for the $\mathcal{A}$-constrained channel $\Phi$ is
nonempty then it maps by the channel $\Phi$ into a single state.

\textbf{Corollary 1.} \textit{If there exists input optimal
average state $\bar{\rho}$ for the $\mathcal{A}$-constrained
channel $\Phi$ then $\Phi(\bar{\rho})=\Omega(\Phi,\mathcal{A})$.}

Note that compactness of the set $\mathcal{A}$ guarantees
existence of at least one input average state.

This corollary justifies the following definition.

\textbf{Definition 1'.} \textit{The state
$\Omega(\Phi,\mathcal{A})$ is called output optimal average state
for the $\mathcal{A}$-constrained channel $\Phi$.}

There exist examples of constrained channels with finite
$\chi$-capacity but with no input optimal average state, for which
the output optimal average state is explicitly determined and plays
an important role in studying of this channels.

\textbf{Corollary 2.} \textit{Let $\mathcal{A}$ be a convex set.
Then}
\[
\bar{C}(\Phi ;\mathcal{A})\geq \chi _{\Phi }(\rho )+H(\Phi (\rho
)\Vert
\Omega(\Phi,\mathcal{A}))\;\;for\;arbitrary\;state\;\,\rho\,\;in\;
\mathcal{A}.
\]
\textit{Proof.} It is sufficient to consider the case
$\bar{C}(\Phi;\mathcal{A})<+\infty$. Let $\{\pi _{i},\rho _{i}\}$
be an arbitrary ensemble such that $\sum_{i}\pi _{i}\rho _{i}=\rho
\in \mathcal{A}$. By proposition 1
\[
\sum_{i}\pi _{i}H(\Phi (\rho
_{i})\Vert\Omega(\Phi,\mathcal{A}))\leq \bar{C}(\Phi
;\mathcal{A}).
\]
This inequality and Donald's identity
\[
\sum_{i}\pi _{i}H(\Phi
(\rho_{i})\Vert\Omega(\Phi,\mathcal{A}))=\chi_{\Phi }(\{\pi
_{i},\rho _{i}\})+H(\Phi (\rho )\Vert\Omega(\Phi,\mathcal{A}))
\]
complete the proof. $\square$

There exists another approach to the definition of the state
$\Omega(\Phi,\mathcal{A})$. It is possible to show that finiteness
of the $\chi$-capacity of the $\mathcal{A}$-constrained channel
$\Phi$ implies compactness of the set
$\overline{\Phi(\mathcal{A})}$.\footnote{We give this assertion
without proof since it will not be used in the strong
argumentations.} For arbitrary ensemble $\{\mu _{j}, \sigma _{j}\}$
with the average $\bar{\sigma}\in\mathcal{A}$ consider the lower
semicontinuous function $F_{\{\mu _{j}, \sigma
_{j}\}}(\rho')=\sum_{j}\mu _{j}H(\Phi (\sigma _{j})\Vert\,\rho')$ on
the set $\overline{\Phi(\mathcal{A})}$. The function
$F(\rho')=\sup_{\sum_{j}\mu _{j}\sigma _{j}\in\mathcal{A}}F_{\{\mu
_{j}, \sigma _{j}\}}(\rho')$ is also lower semicontinuous on the
compact set $\overline{\Phi(\mathcal{A})}$ and, hence, achieves its
minimum on this set. The following proposition asserts, in
particular, that the state $\Omega(\Phi,\mathcal{A})$ can be defined
as the unique minimal point of the function $F(\rho')$.

\textbf{Proposition 2.} \textit{Let $\mathcal{A}$ be a convex set.
The $\chi$-capacity of the $\mathcal{A}$-constrained channel
$\Phi$ can be expressed as
$$
\bar{C}(\Phi
;\mathcal{A})=\inf_{\rho'\in\overline{\Phi(\mathcal{A})}}\left[\sup_{\sum_{j}\mu
_{j}\sigma _{j}\in\mathcal{A}}\sum_{j}\mu _{j}H(\Phi (\sigma
_{j})\Vert\,\rho')\right].
$$
If $\bar{C}(\Phi;\mathcal{A})<+\infty$ then
$\Omega(\Phi,\mathcal{A})$ is the only state on which the infinum in
the right side is achieved.}

\textit{Proof.} If $\bar{C}(\Phi
;\mathcal{A})<+\infty$ then
$F(\Omega(\Phi,\mathcal{A}))=\bar{C}(\Phi;\mathcal{A})$ due to
proposition 1. Let $\varrho'$ be a state such that
$$
\sup_{\sum_{j}\mu _{j}\sigma _{j}\in\mathcal{A}}\sum_{j}\mu
_{j}H(\Phi (\sigma _{j})\Vert\,\varrho')=F(\varrho')\leq
F(\Omega(\Phi,\mathcal{A}))=\bar{C}(\Phi;\mathcal{A}).
$$
then by proposition 1 $\varrho'=\Omega(\Phi,\mathcal{A})$.

If $\bar{C}(\Phi
;\mathcal{A})=+\infty$ then the right side of the expression in proposition 2 is also equal to $+\infty$.
Indeed, if $\rho'$ is a state in $\mathfrak{S}(\mathcal{H}')$ such
that
$$
\sup_{\sum_{j}\mu _{j}\sigma _{j}\in\mathcal{A}}\sum_{j}\mu
_{j}H(\Phi (\sigma _{j})\Vert\,\rho')<+\infty
$$ then equality
(\ref{D-decomp}) valid for arbitrary approximating sequence of
ensembles $\{\pi_{i}^{k},\rho _{i}^{k}\}$ for the
$\mathcal{A}$-constrained channel $\Phi$ implies $\bar{C}(\Phi
;\mathcal{A})<+\infty$.$\square$

Note that the expression for the $\chi$-capacity in the above
proposition can be considered as a generalization of the "mini-max
formula for $\chi^{*}$" in \cite{Sch-West-1} to the case of an
infinite dimensional constrained channel.

\textit{Remark 1.} Propositions 1-2 and corollaries 1-2 does not
hold without assumption of convexity of the set $\mathcal{A}$. To
show this it is sufficient to consider the noiseless channel
$\Phi=\mathrm{Id}$ and the compact set $\mathcal{A}$, consisting
of two states $\rho_{1}$ and $\rho_{2}$ such that
$H(\rho_{1})=H(\rho_{2})<+\infty$ and
$H(\rho_{1}\|\rho_{2})=+\infty$. In this case $\bar{C}(\Phi
;\mathcal{A})=H(\rho_{1})=H(\rho_{2})$, the states $\rho_{1}$ and
$\rho_{2}$ are input optimal average states in the sense of
definition 1 with the different images $\Phi(\rho_{1})=\rho_{1}$
and $\Phi(\rho_{2})=\rho_{2}$.

\section{The $\chi$-function}

The function $\chi_{\Phi}(\rho)$ on $\mathfrak{S}(\mathcal{H})$ is
defined by (\ref{chi-fun-def}). It is shown in \cite{H-Sh-2} that
\begin{equation}\label{chi-def-2}
\chi_{\Phi}(\rho)=\sup_{\pi\in\mathcal{P}_{\{\rho\}}}\int\limits_{\mathfrak{S}(
\mathcal{H})}H(\Phi(\sigma)\Vert\Phi(\rho))\pi(d\sigma),
\end{equation}
where $\mathcal{P}_{\{\rho\}}$ is the set of all probability
measures on $\mathfrak{S}(\mathcal{H})$ with the barycenter
$\rho$, and that under the condition $H(\Phi(\rho))<+\infty$ the
supremum in (\ref{chi-def-2}) is achieved on some measure
supported by pure states.

Note that $H(\Phi(\rho))=+\infty$ does not imply
$\chi_{\Phi}(\rho)=+\infty$. Indeed, it is easy to construct a
channel $\Phi$ from a finite dimensional system into infinite
dimensional one such that $H(\Phi(\rho))=+\infty$ for any
$\rho\in\mathfrak{S}(\mathcal{H})$.\footnote{For example, the
channel
$\Phi:\rho\mapsto\frac{1}{2}\rho\oplus\frac{1}{2}\mathrm{Tr}(\rho)\tau$,
where $\tau$ is a fixed state with infinite entropy.} On the other
hand, by the monotonicity property of the relative entropy
\cite{L-2}
$$\sum_{i}\pi_{i}H(\Phi (\rho
_{i})\|\Phi(\rho))\leq\sum_{i}\pi_{i}H(\rho
_{i}\|\rho)\leq\log\dim\mathcal{H}<+\infty
$$
for arbitrary ensemble $\{\pi_{i},\rho_{i}\}$, and hence
$\chi_{\Phi}(\rho)\leq\log\dim\mathcal{H}<+\infty$ for any
$\rho\in\mathfrak{S}(\mathcal{H})$.

For arbitrary state $\rho$ such that $H(\Phi(\rho))<+\infty$ the
$\chi$-function has the following representation
\begin{equation}\label{chi-exp-1}
\chi _{\Phi}(\rho)=H(\Phi(\rho))-\hat{H}_{\Phi}(\rho),
\end{equation}
where
\begin{equation}\label{conv-def}
\hat{H}_{\Phi}(\rho)=
\inf_{\pi\in\mathcal{P}_{\{\rho\}}}\int\limits_{\mathfrak{S}(\mathcal{H})}H(\Phi(\rho))\pi(d\rho)
=\inf_{\sum_{i}\pi_{i}\rho_{i}=\rho}\sum_{i}\pi_{i}H(\Phi(\rho_{i}))
\end{equation}
is a convex closure of the output entropy $H(\Phi(\rho))$ (this is
proved in \cite{Sh-2}).\footnote{Note that the second equality in
(\ref{conv-def}) holds under the condition
$H(\Phi(\rho))<+\infty$, it is not valid in general (see lemma 2
in \cite{Sh-2} and the notes below).}

Note that the notion of the convex closure of the output entropy
is widely used in the quantum information theory in connection
with the notion of the entanglement of formation (EoF). Namely, in
the finite dimensional case EoF was defined in \cite{B&Ko} as the
convex hull (=convex closure) of the output entropy of a partial
trace channel from the state space of a bipartite system onto the
state space of its single subsystem. In the infinite dimensional
case the definition of EoF as the $\sigma$-convex hull of the
output entropy of a partial trace channel is proposed in
\cite{ESP} while some advantages of the definition of EoF as the
convex closure of the output entropy of a partial trace channel
are considered in \cite{Sh-2}. It is shown that the two above
definitions coincides on the set of states with finite entropy of
partial trace, but their coincidence for arbitrary state remains
an open problem.

In the finite dimensional case the output entropy $H(\Phi(\rho))$
and its convex closure (=convex hull) $\hat{H}_{\Phi}(\rho)$ are
continuous concave and convex functions on
$\mathfrak{S}(\mathcal{H})$ correspondingly and the representation
(\ref{chi-exp-1}) is valid for all states. It follows that in this
case the function $\chi_{\Phi}(\rho)$ is continuous and concave on
$\mathfrak{S}(\mathcal{H})$.

In the infinite dimensional case the output entropy
$H(\Phi(\rho))$ is only lower semicontinuous and, hence, the
function $\chi_{\Phi}(\rho)$ is not continuous even in the case of
the noiseless channel $\Phi$, for which
$\chi_{\Phi}(\rho)=H(\Phi(\rho))$. But it turns out that the
function $\chi_{\Phi}(\rho)$ for arbitrary channel $\Phi$ has
properties similar to the properties of the output entropy
$H(\Phi(\rho))$.

\textbf{Proposition 3.} \textit{The function $\chi_{\Phi}(\rho)$
is a nonnegative concave and lower semicontinuous function on
$\mathfrak{S}(\mathcal{H})$ such that
\begin{equation}\label{chi-concavity}
\chi_{\Phi}(\bar{\rho})-
\sum_{i=1}^{n}\pi_{i}\chi_{\Phi}(\rho_{i})\geq
\sum_{i=1}^{n}\pi_{i}H\left(\Phi(\bar{\rho}_{i})\|\Phi(\bar{\rho})\right)
\end{equation}
for arbitrary ensemble $\{\pi_{i},\rho_{i}\}_{i=1}^{n}$ with the
average state $\bar{\rho}$.}\footnote{This inequality can be
considered as a generalization to the case of the $\chi$-function
of the following well known identity for quantum entropy
$$H(\bar{\rho})- \sum_{i=1}^{n}\pi_{i}H(\rho_{i})=
\sum_{i=1}^{n}\pi_{i}H\left(\rho_{i}\|\bar{\rho}\right).
$$}

\textit{If the restriction of the output entropy $H(\Phi(\rho))$
to a particular subset
$\mathcal{A}\subseteq\mathfrak{S}(\mathcal{H})$ is continuous then
the restriction of the function $\chi_{\Phi}(\rho)$ to this subset
$\mathcal{A}$ is continuous as well.}

\textit{Proof.} Nonnegativity of the $\chi$-function is obvious.
Let us show first its concavity. Note that for a convex set of
states with finite output entropy this concavity easily follows
from (\ref{chi-exp-1}). But to prove concavity on the whole state
space we will use the approach based on lemma 1 and providing
inequality (\ref{chi-concavity}). Let $\varepsilon>0$ be
arbitrary. By definition of the $\chi$-function for each
$i=\overline{1,n}\,$ there exists ensemble
$\{\mu^{i}_{j},\sigma^{i}_{j}\}_{j=1}^{m(i)}$ with the average
$\rho_{i}$ such that
$\chi_{\Phi}(\{\mu^{i}_{j},\sigma^{i}_{j}\})>\chi_{\Phi}(\rho_{i})-\varepsilon$.
Since the average state of the ensemble
$\sum_{i=1}^{n}\pi_{i}\{\mu^{i}_{j},\sigma^{i}_{j}\}$ coincides
with $\bar{\rho}$, by using lemma 1 we have
$$
\begin{array}{c}
\chi_{\Phi}(\bar{\rho})\geq\chi_{\Phi}\left(\sum_{i=1}^{n}\pi_{i}\{\mu^{i}_{j},\sigma^{i}_{j}\}\right)\geq
\sum_{i=1}^{n}\pi_{i}\chi_{\Phi}(\{\mu^{i}_{j},\sigma^{i}_{j}\})\\\\
+\sum_{i=1}^{n}\pi_{i}H\left(\Phi(\bar{\rho}_{i})\|\Phi(\bar{\rho})\right)\geq
\sum_{i=1}^{n}\pi_{i}\chi_{\Phi}(\rho_{i})
+\sum_{i=1}^{n}\pi_{i}H\left(\Phi(\bar{\rho}_{i})\|\Phi(\bar{\rho})\right)-\varepsilon.
\end{array}
$$
Since $\varepsilon$ can be arbitrary small inequality
(\ref{chi-concavity}) is established. It obviously implies
concavity of the $\chi$-function.

To prove lower semicontinuity of the $\chi$-function we have to
show
\begin{equation}  \label{chi-l-s-c}
\liminf_{n\rightarrow+\infty}\chi_{\Phi}(\rho_{n})\geq\chi _{\Phi
}(\rho_{0}).
\end{equation}
for arbitrary state $\rho_{0}$ and arbitrary sequence $\rho_{n}$
converging to this state $\rho_{0}$.

For arbitrary $\varepsilon>0$ let $\{\pi_{i},\rho_{i}\}$ be an
ensemble with the average $\rho_{0}$ such that
\[
\sum_{i}\pi_{i}H(\Phi(\rho_{i})\|\Phi(\rho_{0}))\geq
\chi_{\Phi}(\rho_{0})-\varepsilon.
\]
By lemma 3 below there exists the sequence of ensembles
$\{\pi_{i}^{n}, \rho_{i}^{n}\}$ of fixed size such that
\[
\lim_{n\rightarrow+\infty}\pi^{n}_{i}=\pi_{i},\quad
\lim_{n\rightarrow+\infty}\rho^{n}_{i}=\rho_{i},\quad\mathrm{and}\quad
\rho_{n}=\sum_{i=1}^{m}\pi^{n}_{i}\rho^{n}_{i}.
\]

By definition we have
\[
\begin{array}{c}
 \liminf_{n\rightarrow+\infty}\chi_{\Phi}(\rho_{n})\geq
\liminf_{n\rightarrow+\infty}\sum_{i}\pi_{i}^{n}H(\Phi(\rho_{i}^{n})\|\Phi(\rho_{n}))
\\\\\geq \sum_{i}\pi_{i}H(\Phi(\rho_{i})\|\Phi(\rho_{0}))\geq
\chi_{\Phi}(\rho_{0})-\varepsilon,
\end{array}
\]
where lower semicontinuity of the relative entropy was used. This
implies (\ref{chi-l-s-c}) (due to the freedom of the choice of
$\varepsilon$).

The last assertion of proposition 3 follows from the representation
(\ref{chi-exp-1}) and from lower semicontinuity of the function
$\hat{H}_{\Phi}(\rho)$ established in \cite{Sh-2}. $\square$

\textbf{Lemma 3.} \textit{Let $\{\pi_{i},\rho_{i}\}$ be an
arbitrary ensemble of $m$ states with the average state $\rho$ and
let $\{\rho_{n}\}$ be an arbitrary sequence of states converging
to the state $\rho$. There exists the sequence
$\{\pi^{n}_{i},\rho^{n}_{i}\}$ of ensembles of $m$ states such
that
\[
\lim_{n\rightarrow+\infty}\pi^{n}_{i}=\pi_{i},\quad
\lim_{n\rightarrow+\infty}\rho^{n}_{i}=\rho_{i},\quad and \quad
\rho_{n}=\sum_{i=1}^{m}\pi^{n}_{i}\rho^{n}_{i}.
\]}

\textit{Proof.} Without loss of generality we may assume that $\pi
_{i}>0$ for all $i$. Let $\mathcal{D}\subseteq \mathcal{H}$ be the
support of $\rho =\sum_{i=1}^{m}\pi _{i}\rho _{i}$ and $P$ be the
projector onto $\mathcal{D}$. Since $\rho _{i}\leq \pi
_{i}^{-1}\rho$ we have
\[
0\leq A_{i}\equiv \rho ^{-1/2}\rho _{i}\rho ^{-1/2}\leq \pi
_{i}^{-1}I,
\]
where we denote by $\rho ^{-1/2}$ the generalized (Moore-Penrose)
inverse of the operator $\rho ^{1/2}$ (equal 0 on the orthogonal
complement to $\mathcal{D}$).

Consider the sequence $B_{i}^{n}=\rho _{n}^{1/2}A_{i}\rho
_{n}^{1/2}+\rho_{n}^{1/2}(I_{\mathcal{H}}-P)\rho_{n}^{1/2}$ of
operators in $\mathfrak{B}(\mathcal{H})$. Since
$\lim_{n\rightarrow +\infty }\rho_{n}=\rho=P\rho$ in the trace
norm, we have
$$
\lim_{n\rightarrow +\infty }B_{i}^{n}=\rho ^{1/2}A_{i}\rho
^{1/2}=\rho _{i}
$$
in the weak operator topology. The last equality implies
$A_{i}\neq 0.$ Note that
$\mathrm{Tr}B_{i}^{n}=\mathrm{Tr}A_{i}\rho
_{n}+\mathrm{Tr}(I_{\mathcal{H}}-P)\rho_{n}<+\infty $ and hence
$$
\lim_{n\rightarrow
+\infty}\mathrm{Tr}B_{i}^{n}=\mathrm{Tr}A_{i}\rho
=\mathrm{Tr}\rho_{i}=1.
$$

Denote by $\rho _{i}^{n}=(\mathrm{Tr}B_{i}^{n})^{-1}B_{i}^{n}$  a
state and by $\pi _{i}^{n}=\pi _{i}\mathrm{Tr}B_{i}^{n}$ a
positive number
for each $i,$ then $\lim_{n\rightarrow +\infty }\pi _{i}^{n}=\pi _{i}$ and $%
\lim_{n\rightarrow +\infty }\rho _{i}^{n}=\rho _{i}$ in the weak
operator topology and hence, by the result in \cite{D-A}, in the
trace norm. Moreover,
$$
\sum_{i=1}^{m}\pi _{i}^{n}\rho _{i}^{n}=\sum_{i=1}^{m}\pi
_{i}B_{i}^{n}=\rho _{n}^{1/2}\rho ^{-1/2}\sum_{i=1}^{m}\pi
_{i}\rho _{i}\rho ^{-1/2}\rho
_{n}^{1/2}+\rho_{n}^{1/2}(I_{\mathcal{H}}-P)\rho_{n}^{1/2}=\rho
_{n}.\quad \square
$$

In the modern convex analysis the notion of strong convexity
(concavity) plays an essential role \cite{P&B}. By using
inequality (\ref{rel-entr-ineq}) and proposition 3 we obtain the
following observation.

\textbf{Corollary 3.} \textit{$\chi_{\Phi}(\rho)$ is a strongly
concave function on $\mathfrak{S}(\mathcal{H})$ in the following
sense
$$
\chi_{\Phi}(\lambda\rho_{1}+(1-\lambda)\rho_{2})\geq
\lambda\chi_{\Phi}(\rho_{1})+(1-\lambda)\chi_{\Phi}(\rho_{2})+
\textstyle\frac{1}{2}\lambda(1-\lambda)\|\Phi(\bar{\rho}_{2})-\Phi(\bar{\rho}_{1})\|_{1}^{2}.
$$
for arbitrary $\rho_{1}$ and $\rho_{2}$ in
$\mathfrak{S}(\mathcal{H})$.}

The similarity of the properties of the functions
$\chi_{\Phi}(\rho)$ and $H(\Phi(\rho))$ is stressed by the
following analog of Simon's dominated convergence theorem for
quantum entropy \cite{Simon}, which will be used later.

\textbf{Corollary 4.} \textit{Let $\rho_{n}$ be a sequence of
states in $\mathfrak{S}(\mathcal{H})$, converging to the state
$\rho$ and such that $\lambda_{n}\rho_{n}\leq\rho$ for some
sequence $\lambda_{n}$ of positive numbers, converging to $1$.
Then
$$
\lim_{n\rightarrow+\infty}\chi_{\Phi}(\rho_{n})=\chi_{\Phi}(\rho).
$$}
\textit{Proof.} The condition $\lambda_{n}\rho_{n}\leq\rho$
implies decomposition
$\rho=\lambda_{n}\rho_{n}+(1-\lambda_{n})\rho'_{n}$, where
$\rho'_{n}=(1-\lambda_{n})^{-1}(\rho-\lambda_{n}\rho_{n})$ is a
state. By concavity of the $\chi$-function we have
$$
\chi_{\Phi}(\rho)\geq\lambda_{n}\chi_{\Phi}(\rho_{n})+
(1-\lambda_{n})\chi_{\Phi}(\rho'_{n})\geq
\lambda_{n}\chi_{\Phi}(\rho_{n}),
$$
which implies
$\limsup_{n\rightarrow+\infty}\chi_{\Phi}(\rho_{n})\leq\chi_{\Phi}(\rho)$.
This and lower semicontinuity of the $\chi$-function completes the
proof.$\square$

\textit{Example 1.} Let $H^{\prime}$ be a positive unbounded
operator on the space $\mathcal{H}^{\prime }$ such that
$\mathrm{Tr}\exp (-\beta H^{\prime })<+\infty$  for all $\beta >0$
and $h^{\prime}$ be a positive number. In the proof of proposition
3 in \cite{H-Sh-2} continuity of the restriction of the output
entropy $H(\Phi(\rho))$ to the subset
$\mathcal{A}_{h^{\prime}}=\{\rho\in\mathfrak{S}(\mathcal{H})\,|\,\mathrm{Tr}\,\Phi
(\rho)H^{\prime }\leq h^{\prime}\}$ was established.\footnote{The
value $\mathrm{Tr}\,\Phi (\rho)H^{\prime }$ is defined as a limit
of nondecreasing sequence
$\mathrm{Tr}\,\Phi(\rho)Q^{\prime}_{n}H^{\prime }$, where
$Q^{\prime }_{n}$ is the spectral projector of $H^{\prime}$
corresponding to the lowest $n$ eigenvalues \cite{H-c-w-c}.} By
proposition 3 the restriction of the $\chi$-function to the set
$\mathcal{A}_{h^{\prime}}$ is continuous. As it is mentioned in
\cite{H-Sh-2}, the above continuity condition is fulfilled for
Gaussian channels with the power constraint of the form
$\mathrm{Tr}\rho H\leq h$, where $H=R^{T}\epsilon R$ is the
many-mode oscillator Hamiltonian with nondegenerate energy matrix
$\epsilon$ and $R$ are the canonical variables of the system.

We shall use the following chain properties of the
$\chi$-function.

\textbf{Proposition 4.} \textit{Let
$\Phi:\mathfrak{S}(\mathcal{H})\mapsto\mathfrak{S}(\mathcal{H}')$
and
$\Psi:\mathfrak{S}(\mathcal{H}')\mapsto\mathfrak{S}(\mathcal{H}'')$
be two channels. Then}
$$
\chi_{\Psi\circ\Phi}(\rho)\leq\chi_{\Phi}(\rho)\quad and \quad
\chi_{\Psi\circ\Phi}(\rho)\leq\chi_{\Psi}(\Phi(\rho))
\;\;\;for\;arbitrary\;\,\rho\,\;in\;\,\mathfrak{S}(\mathcal{H}).
$$

\textit{Proof.} The first inequality follows from the monotonicity
property of the relative entropy \cite{L-2} and
(\ref{chi-fun-def}), while the second one is a direct corollary of
the definition (\ref{chi-fun-def}) of the
$\chi$-function.$\square$

\section{On continuity of the $\chi$-capacity}

In this section the question of continuity of the $\chi$-capacity as
a function of channel is considered. Dealing with this question we
must choose a topology on the set
$\mathcal{C}(\mathcal{H},\mathcal{H}')$ of all quantum channels from
$\mathfrak{S}(\mathcal{H})$ into $\mathfrak{S}(\mathcal{H}')$. This
choice is essential only in the infinite dimensional case because
all locally convex Hausdorff topologies on a finite dimensional
space are equivalent.

Let $\mathcal{L}(\mathcal{H},\mathcal{H}')$ be the linear space of
all continuous linear mapping from $\mathfrak{T}(\mathcal{H})$
into $\mathfrak{T}(\mathcal{H}')$. We will use the topology on
$\mathcal{C}(\mathcal{H},\mathcal{H}')\subset\mathcal{L}(\mathcal{H},\mathcal{H}')$
generated by the topology of strong convergence on
$\mathcal{L}(\mathcal{H},\mathcal{H}')$.

\textbf{Definition 2.} \textit{The topology on the linear space
$\mathcal{L}(\mathcal{H},\mathcal{H}')$ defined by the family of
seminorms
$\{\|\Phi\|_{\rho}=\|\Phi(\rho)\|_{1}\}_{\rho\in\mathfrak{T}(\mathcal{H})}$
is called the topology of strong convergence.}

Since an arbitrary operator in $\mathfrak{T}(\mathcal{H})$ can be
represented as a linear combination of operators in
$\mathfrak{S}(\mathcal{H})$ it is possible to consider only
seminorms $\|\cdot\|_{\rho}$ corresponding to
$\rho\in\mathfrak{S}(\mathcal{H})$ in the above definition.

Note that  a sequence $\Phi_{n}$ of channels in
$\mathcal{C}(\mathcal{H},\mathcal{H}')$ strongly converges to a
channel $\Phi\in\mathcal{C}(\mathcal{H},\mathcal{H}')$ if and only
if $\lim_{n\rightarrow+\infty}\Phi_{n}(\rho)=\Phi(\rho)$ for all
$\rho\in\mathfrak{S}(\mathcal{H})$. Due to the result in
\cite{D-A} the above limit may be in the weak operator topology.

\textbf{Theorem 1.} \textit{Let $\mathcal{A}$ be an arbitrary
closed and convex subset of
$\mathfrak{S}(\mathcal{H})$.}\footnote{Convexity of $\mathcal{A}$
is used only in the proof of (\ref{a-s-l}).}

\textit{In the case of finite dimensional spaces $\mathcal{H}$ and
$\mathcal{H}'$ the $\chi$-capacity $\bar{C}(\Phi,\mathcal{A})$ is
a continuous function on the set
$\mathcal{C}(\mathcal{H},\mathcal{H}')$. If $\Phi_{n}$ is an
arbitrary sequence of channels in
$\mathcal{C}(\mathcal{H},\mathcal{H}')$, converging to some
channel $\Phi$ in $\mathcal{C}(\mathcal{H},\mathcal{H}')$, then
there exists}
\begin{equation}\label{a-s-l}
\lim_{n\rightarrow\infty}\Omega(\Phi_{n},\mathcal{A})=\Omega(\Phi,\mathcal{A}).
\end{equation}

\textit{In general the $\chi$-capacity $\bar{C}(\Phi,\mathcal{A})$
is a lower semicontinuous function on the set
$\mathcal{C}(\mathcal{H},\mathcal{H}')$ equipped with the topology
of strong convergence.}

\textit{Proof.} Let us first show lower semicontinuity of the
$\chi$-capacity. Let $\varepsilon>0$ and $\Phi_{\lambda}$ be an
arbitrary net of channels, strongly converging to the channel
$\Phi$, and $\{\pi_{i},\rho_{i}\}$ be an  ensemble with the
average $\bar{\rho}$ such that
$\chi_{\Phi}(\{\pi_{i},\rho_{i}\})>\bar{C}(\Phi,
\mathcal{A})-\varepsilon$.  By lower semicontinuity of the
relative entropy \cite{W}
$$\liminf_{\lambda}\sum_{i}\pi_{i}H(\Phi_{\lambda}(\rho
_{i})\|\Phi_{\lambda}(\bar{\rho}))\geq \sum_{i}\pi_{i}H(\Phi (\rho
_{i})\|\Phi(\bar{\rho}))>\bar{C}(\Phi,\mathcal{A})-\varepsilon
$$
This implies
\[
\liminf_{\lambda}\bar{C}(\Phi_{\lambda},\mathcal{A})\geq
\bar{C}(\Phi,\mathcal{A}).
\]
It follows that
\begin{equation}\label{l-s-c}
\liminf_{n\rightarrow+\infty}\bar{C}(\Phi_{n},\mathcal{A})\geq
\bar{C}(\Phi,\mathcal{A})
\end{equation}
for arbitrary sequence $\Phi_{n}$ of channels strongly converging
to a channel $\Phi$.

Now to prove the continuity of the $\chi$-capacity in the finite
dimensional case it is sufficient to show that for the above
sequence of channels
\begin{equation}\label{u-s-c}
\limsup_{n\rightarrow+\infty}\bar{C}(\Phi_{n},\mathcal{A})\leq
\bar{C}(\Phi,\mathcal{A}).
\end{equation}
For an arbitrary $\mathcal{A}$-constrained channel from
$\mathcal{C}(\mathcal{H},\mathcal{H}')$ there exists optimal
ensemble consisting of $m=(\dim \mathcal{H})^{2}$ states
(probably, some states with zero weights)
\cite{Dev-1},\cite{Sch-West-1}. Let $\mathfrak{P}$ be the compact
space of all probability distributions with $m$ outcomes. Consider
the compact space\footnote{with product topology}
$$
\mathfrak{PC}^{m}=\mathfrak{P}\times\underbrace{\mathfrak{S}(\mathcal{H})\times...
\times\mathfrak{S}(\mathcal{H})}_{m},
$$
consisting of sequences $(\{\pi_{i}\}_{i=1}^{m},
\rho_{1},...,\rho_{m})$, corresponding to arbitrary input ensemble
$\{\pi_{i},\rho_{i}\}_{i=1}^{m}$ of $m$ states.

Suppose (\ref{u-s-c}) is not true. Without loss of generality we
may assume that
\begin{equation}\label{u-s-c-contr}
\lim_{n\rightarrow+\infty}\bar{C}(\Phi_{n},\mathcal{A})>
\bar{C}(\Phi,\mathcal{A}).
\end{equation}
Let $\{\pi_{i}^{n},\rho_{i}^{n}\}_{i=1}^{m}$ be an optimal
ensemble for the $\mathcal{A}$-constrained channel $\Phi_{n}$. By
compactness of $\mathfrak{PC}^{m}$ we can choose a subsequence
$(\{\pi_{i}^{n_{k}}\}_{i=1}^{m},
\rho_{1}^{n_{k}},...,\rho_{m}^{n_{k}})$ converging to some element
$(\{\pi_{i}^{*}\}_{i=1}^{m}, \rho_{1}^{*},...,\rho_{m}^{*})$ of
the space $\mathfrak{PC}^{m}$. By definition of the product
topology on $\mathfrak{PC}^{m}$ it means that
$$
\lim_{k\rightarrow\infty}\pi_{i}^{n_{k}}=\pi_{i}^{*},\quad
\lim_{k\rightarrow\infty}\rho_{i}^{n_{k}}=\rho_{i}^{*}.
$$
The average state of the ensemble
$\{\pi_{i}^{*},\rho_{i}^{*}\}_{i=1}^{m}$ is a limit of the
sequence of average states of the ensembles
$\{\pi_{i}^{n_{k}},\rho_{i}^{n_{k}}\}_{i=1}^{m}$ and hence lies in
$\mathcal{A}$ (which is closed by the assumption).

By continuity of the quantum entropy in finite dimensional case we
have
$$
 \lim_{k\rightarrow+\infty}\bar{C}(\Phi_{n_{k}}, \mathcal{A})=
\lim_{k\rightarrow+\infty}\chi_{\Phi_{n_{k}}}(\{\pi_{i}^{n_{k}},\rho_{i}^{n_{k}}\})=
\chi_{\Phi}(\{\pi_{i}^{*},\rho_{i}^{*}\})\leq\bar{C}(\Phi,
\mathcal{A}),
$$
which contradicts to (\ref{u-s-c-contr}).

Comparing (\ref{l-s-c}) and (\ref{u-s-c}) we see that
$$
\lim_{n\rightarrow+\infty}\bar{C}(\Phi_{n},
\mathcal{A})=\bar{C}(\Phi, \mathcal{A}).
$$
It follows that the above ensemble
$\{\pi_{i}^{*},\rho_{i}^{*}\}_{i=1}^{m}$ is optimal for the
$\mathcal{A}$-constrained channel $\Phi$. Hence, there exists the
input optimal average state $\bar{\rho}^{*}$ for the
$\mathcal{A}$-constrained channel $\Phi$ which is a partial limit
of the sequence $\{\bar{\rho}^{n}\}$ of the input optimal average
states for the $\mathcal{A}$-constrained channels $\Phi_{n}$.

Suppose (\ref{a-s-l}) is not true. Without loss of generality we
may (by compactness argument) assume that there exists
$\;\lim_{n\rightarrow\infty}\Omega(\Phi_{n},\mathcal{A})\neq\Omega(\Phi,\mathcal{A})\;$.
By proposition 1 this contradicts to the previous observation.
$\square$

The assumption of finite dimensionality in the first part of
theorem 1 is essential. The following example shows that generally
the $\chi$-capacity is not continuous function of a channel even
in the stronger trace norm topology on the space of all channels.
The example is a purely classical channel which has a standard
extension to a quantum one.

\textit{Example 2.} Consider Abelian von Neumann algebra
$\mathbf{\textit{l}}_{\infty}$ and its predual
$\mathbf{\textit{l}}_{1}$. Let $\{\Phi^{q}_{n}; n=1,2,...;
q\in(0,1)\}$ be the family of classical unconstrained channels
defined by the formula
$$
\begin{array}{c}
\Phi^{q}_{n}(\{x_{1},x_{2},...,x_{n},...\})=\{(1-q)\sum_{i=1}^{\infty}x_{i},
q\sum_{i=n+1}^{\infty}x_{i},qx _{1},...,qx_{n}, 0, 0, ...\}
\end{array}
$$
for $\{x_{1},x_{2},...,x_{n},...\}\!\in\!\mathbf{\textit{l}}_{1}$.
Defining
$
\Phi^{0}(\{x_{1},x_{2},...,x_{n},...\})\!=\!\{\sum_{i=1}^{\infty}x_{i},
0,0,...\}
$
we have
$$
\begin{array}{c}
\|(\Phi^{q}_{n}-\Phi^{0})(\{x_{i}\}_{i=1}^{\infty})\|_{1}=q\|\{-\sum_{i=1}^{\infty}x_{i},
\sum_{i=n+1}^{\infty}x_{i}, x_{1},..., x_{n}, 0, 0,
...\}\|_{1}\\\\=q(|\sum_{i=1}^{\infty}x_{i}|+
|\sum_{i=n+1}^{\infty}x_{i}|+|x_{1}|+...+|x_{n}|)\leq
3q\|\{x_{i}\}_{i=1}^{\infty}\|_{1},
\end{array}
$$
hence $\|\Phi^{q}_{n}-\Phi^{0}\|\rightarrow 0$ as $q\rightarrow 0$
uniformly in $n$.

To evaluate the $\chi$-capacity of the channel $\Phi^{q}_{n}$ it
is sufficient to note that
$H(\Phi^{q}_{n}(\mathrm{any\;pure\;state}))=h_{2}(q)=-q\log
q-(1-q)\log(1-q)$ and
$$
 H(\Phi^{q}_{n}(\mathrm{any\;state}))\leq
H(\Phi^{q}_{n}(\{\underbrace{\textstyle\frac{1}{n+1},\frac{1}{n+1},...,\frac{1}{n+1}}_{n+1},
0, 0 ...\}))=q\log(n+1)+h_{2}(q).
$$
It follows by definition that
$\bar{C}(\Phi^{q}_{n})=q\log(n+1),\;q\in(0,1),\; n\in \mathbb{N}$.

Take arbitrary $C$ such that $0<C\leq+\infty$ and choose a
sequence $q(n)$ such that $\,\lim_{n\rightarrow\infty}q(n)=0\,$
while $\,\lim_{n\rightarrow\infty}q(n)\log(n+1)=C\,$. Then we have
$\;\lim_{n\rightarrow\infty}\|\Phi^{q(n)}_{n}-\Phi^{0}\|=0\;$ but
$\;\lim_{n\rightarrow\infty}\bar{C}(\Phi^{q(n)}_{n})=C>0=\bar{C}(\Phi^{0})$.$\square$

\textit{Remark 2.} The above example demonstrates harsh
discontinuity of the $\chi$-capacity in the infinite dimensional
case. One can see that a similar discontinuity underlies Shor's
construction \cite{Sh-e-a-q} allowing to prove equivalence of
different additivity properties by using channel extension and a
limiting procedure.

\section{Additivity for constrained channels}

Let $\Phi :\mathfrak{S}(\mathcal{H})\mapsto
\mathfrak{S}(\mathcal{H}^{\prime })$ and $\Psi
:\mathfrak{S}(\mathcal{K})\mapsto \mathfrak{S}(\mathcal{K}^{\prime
})$ be two channels with the constraints, defined by closed
subsets $ \mathcal{A}\subset \mathfrak{S}(\mathcal{H})$ and
$\mathcal{B}\subset \mathfrak{S}(\mathcal{K})$ correspondingly.
For the channel $\Phi\otimes \Psi$ we consider the constraint
defined by the requirements $\bar{\omega}^{\mathcal{H}
}:=\mathrm{Tr}_{\mathcal{K}}\bar{\omega}\in \mathcal{A}$ and
$\bar{\omega}^{\mathcal{K}}:=\mathrm{Tr}_{\mathcal{H}
}\bar{\omega}\in \mathcal{B}$, where $\bar{\omega}$ is the average
state of an input ensemble $\{\mu _{i},\omega _{i}\}$. The subset
of $\mathfrak{S}(\mathcal{H}\otimes \mathcal{K})$ consisting of
states $\omega$ such that $\mathrm{Tr}_{\mathcal{K}}\omega\in
\mathcal{A}$ and $\mathrm{Tr}_{\mathcal{H}}\omega\in \mathcal{B}$
will be denoted $\mathcal{A}\otimes \mathcal{B}$.

\textbf{Lemma 4.} \textit{The set $\mathcal{A}\otimes \mathcal{B}$
is convex subset of $\mathfrak{S}(\mathcal{H}\otimes \mathcal{K})$
if and only if the sets $\mathcal{A}$ and $\mathcal{B}$ are convex
subsets of $\mathfrak{S}(\mathcal{H})$ and of
$\mathfrak{S}(\mathcal{K})$ correspondingly.}

\textit{The set $\mathcal{A}\otimes \mathcal{B}$ is compact subset
of $\mathfrak{S}(\mathcal{H}\otimes \mathcal{K})$ if and only if
the sets $\mathcal{A}$ and $\mathcal{B}$ are compact subsets of
$\mathfrak{S}(\mathcal{H})$ and of $\mathfrak{S}(\mathcal{K})$
correspondingly.}

\textit{Proof.} The first statement of this lemma is trivial. To
prove the second note that compactness of the set
$\mathcal{A}\otimes \mathcal{B}$ implies compactness of the sets
$\mathcal{A}$ and $\mathcal{B}$ due to continuity of  partial
trace.

The proof of the converse implication is based on the following
characterization of a compact set  of states : \textit{a closed
subset $\mathcal{A}$ of $\mathfrak{S}(\mathcal{H})$ is compact if
and only if for any $\varepsilon>0$ there exists finite
dimensional projector $P_{\varepsilon}$ such that
$\mathrm{Tr}P_{\varepsilon}\rho>1-\varepsilon$ for all
$\rho\in\mathcal{A}$}. This characterization can be deduced by
combining results of \cite{S} and \cite{D-A} (see the proof of the
lemma in \cite{H-c-w-c}). Its proof is also presented in the
Appendix of \cite{H-Sh-2}.

Let $\mathcal{A}$ and $\mathcal{B}$ be compact. By the above
characterization for arbitrary $\varepsilon>0$ there exist finite
rank projectors $P_{\varepsilon}$ and $Q_{\varepsilon}$ such that
$$
\mathrm{Tr}P_{\varepsilon}\rho>1-\varepsilon,\;\forall\rho\in\mathcal{A}\quad
\mathrm{and}\quad
\mathrm{Tr}Q_{\varepsilon}\sigma>1-\varepsilon,\;\forall\sigma\in\mathcal{B}.
$$
Since $\omega^{\mathcal{H}}\in\mathcal{A}$ and
$\omega^{\mathcal{K}}\in\mathcal{B}$ for arbitrary
$\omega\in\mathcal{A}\otimes \mathcal{B}$ we have
$$
\begin{array}{c}
 \mathrm{Tr}((P_{\varepsilon}\otimes Q_{\varepsilon})\cdot\omega)=\mathrm{Tr}((P_{\varepsilon}\otimes
I_{\mathcal{K}})\cdot\omega)-\mathrm{Tr}(P_{\varepsilon}\otimes
(I_{\mathcal{K}}-Q_{\varepsilon}))\cdot\omega)\\\\\geq
\mathrm{Tr}P_{\varepsilon}\omega^{\mathcal{H}}-
\mathrm{Tr}(I_{\mathcal{K}}-Q_{\varepsilon})\omega^{\mathcal{K}}>1-2\varepsilon.
\end{array}
$$
The above characterization implies compactness of the set
$\mathcal{A}\otimes \mathcal{B}$.$\square$

The conjecture of additivity of the $\chi$-capacity for the
$\mathcal{A}$-constrained channel $\Phi$ and the
$\mathcal{B}$-constrained channel $\Psi$ is
\cite{H-Sh},\cite{H-Sh-2}
\begin{equation}
\bar{C}\left( \Phi \otimes \Psi ;\mathcal{A}\otimes
\mathcal{B}\right) =\bar{ C}(\Phi ;\mathcal{A})+\bar{C}(\Psi
;\mathcal{B}).  \label{addit}
\end{equation}

\textit{Remark 3.} Let $\Omega(\Phi;\mathcal{A})$ and
$\Omega(\Psi;\mathcal{B})$ be the output optimal average states
for the $\mathcal{A}$-constrained channel $\Phi$ and the
$\mathcal{B}$-constrained channel $\Psi$ correspondingly.
Additivity of the $\chi$-capacity (\ref{addit}) implies that
$\Omega(\Phi;\mathcal{A})\otimes\Omega(\Psi;\mathcal{B})$ is the
output optimal average state for the
$\mathcal{A}\otimes\mathcal{B}$-constrained channel
$\Phi\otimes\Psi$. Indeed, let $\{\pi^{k}_{i};\rho^{k}_{i}\}$ and
$\{\mu^{k}_{j};\sigma^{k}_{j}\}$ be approximating sequences of
ensembles for the $\mathcal{A}$-constrained channel $\Phi$ and the
$\mathcal{B}$-constrained channel $\Psi$. By proposition 1 the
sequences $\{\Phi(\bar{\rho}^{k})\}$ and
$\{\Psi(\bar{\sigma}^{k})\}$ converge to
$\Omega(\Phi;\mathcal{A})$ and to $\Omega(\Psi;\mathcal{B})$
correspondingly. By (\ref{addit}) the sequence of ensembles
$\{\pi^{k}_{i}\mu^{k}_{j};\rho^{k}_{i}\otimes\sigma^{k}_{j}\}$ is
an approximating sequence for the $\mathcal{A}\otimes
\mathcal{B}$-constrained channel $\Phi\otimes\Psi$. By proposition
1 the limit
$\Omega(\Phi;\mathcal{A})\otimes\Omega(\Psi;\mathcal{B})$ of the
sequence $\{\Phi(\bar{\rho}^{k})\otimes\Psi(\bar{\sigma}^{k})\}$
is the output optimal average states for the
$\mathcal{A}\otimes\mathcal{B}$-constrained channel
$\Phi\otimes\Psi$.

The results of the previous sections make possible to obtain the
following infinite dimensional version of theorem 1 in
\cite{H-Sh}.

\textbf{Theorem 2.} \textit{Let $\Phi
:\mathfrak{S}(\mathcal{H})\mapsto \mathfrak{S}(\mathcal{H}^{\prime
})$ and $\Psi
:\mathfrak{S}(\mathcal{K})\mapsto \mathfrak{S}(\mathcal{K}^{\prime
})$ be arbitrary channels. The following properties are
equivalent:}

$\mathit{(i)}$ \textit{equality (\ref{addit}) holds for arbitrary
subsets $\mathcal{A}\subseteq\mathfrak{S}(\mathcal{H})$ and
$\mathcal{B}\subseteq\mathfrak{S}(\mathcal{K})$ such that
$H(\Phi(\rho))<+\infty$ for all $\rho\in\mathcal{A}$ and
$H(\Psi(\sigma))<+\infty$ for all
$\sigma\in\mathcal{B}$;}\vspace{5pt}

$\mathit{(ii)}$ \textit{inequality
\begin{equation}
\chi _{\Phi \otimes \Psi }(\omega )\leq \chi
_{\Phi}(\omega^{\mathcal{H}})+\chi_{\Psi}(\omega^{\mathcal{K}})
\label{sub-add}
\end{equation}
holds for arbitrary state $\omega$ such that
$H(\Phi(\omega^{\mathcal{H}}))<+\infty$ and
$H(\Psi(\omega^{\mathcal{K}}))<+\infty$;}\vspace{5pt}

$\mathit{(iii)}$ \textit{inequality
\begin{equation}\label{super-add}
\hat{H}_{\Phi\otimes\Psi}(\omega )\geq \hat{H}_{\Phi}
(\omega^{\mathcal{H}} )+ \hat{H}_{\Psi} (\omega^{\mathcal{K}});
\end{equation}
holds for arbitrary state $\omega$ such that
$H(\Phi(\omega^{\mathcal{H}}))<+\infty$ and
$H(\Psi(\omega^{\mathcal{K}}))<+\infty$.}\vspace{5pt}

\textit{Proof.} $(i)\Rightarrow (iii).$ Let $\omega$ be an
arbitrary state with finite $H(\Phi(\omega^{\mathcal{H}}))$ and
$H(\Psi(\omega^{\mathcal{K}}))$.  The validity of $(i)$ implies
\[
\bar{C}\left( \Phi \otimes \Psi ;\{\omega ^{\mathcal{H}}\}\otimes
\{\omega ^{\mathcal{K}}\}\right)=\bar{C}(\Phi ;\{\omega
^{\mathcal{H}}\})+\bar{C}(\Psi
;\{\omega ^{\mathcal{K}}\}).
\]
By remark 3 the state $\Phi(\omega
^{\mathcal{H}})\otimes\Psi(\omega ^{\mathcal{K}})$ is the output
optimal average state for the $\{\omega ^{\mathcal{H}}\}\otimes
\{\omega ^{\mathcal{K}}\}$ -constrained channel $\Phi \otimes \Psi
$.  Noting that $\omega\in\{\omega ^{\mathcal{H}}\}\otimes
\{\omega ^{\mathcal{K}}\}$ and applying corollary 1 we obtain
\begin{equation}\label{opt-c}
\begin{array}{c}
\chi_{\Phi}(\omega ^{\mathcal{H}})+\chi_{\Psi}(\omega
^{\mathcal{K}})=\bar{C}(\Phi
;\{\omega ^{\mathcal{H}}\})+\bar{C}(\Psi;\{\omega ^{\mathcal{K}}\})\\\\
=\bar{C}\left(\Phi\otimes\Psi;\{\omega ^{\mathcal{H}}\}\otimes
\{\omega ^{\mathcal{K}}\}\right)\\\\\geq\chi_{\Phi\otimes
\Psi}(\omega )+H((\Phi\otimes\Psi)(\omega)\Vert\Phi(\omega
^{\mathcal{H}})\otimes \Psi(\omega ^{\mathcal{K}})).
\end{array}
\end{equation}
Due to
$$
H((\Phi \otimes \Psi )(\omega )\Vert \Phi (\omega
^{\mathcal{H}})\otimes \Psi (\omega ^{\mathcal{K}}))=H(\Phi(\omega
^{\mathcal{H}}))+H(\Psi(\omega ^{\mathcal{K}}))-H((\Phi\otimes
\Psi)(\omega))
$$
the inequality (\ref{opt-c}) together with (\ref {chi-exp-1})
implies (\ref{super-add}).

$(iii)\Rightarrow (ii)$. It can be derived from expression (\ref
{chi-exp-1}) for the $\chi$-function and subadditivity of the
(output) entropy.

$(ii)\Rightarrow (i)$. It follows from the definition of the
$\chi$-capacity (\ref{ccap-1}) and inequality (\ref {sub-add})
that
\[
\bar{C}\left( \Phi \otimes \Psi ;\mathcal{A}\otimes
\mathcal{B}\right) \leq \bar{C}(\Phi ;\mathcal{A})+\bar{C}(\Psi
;\mathcal{B}).
\]
Since the converse inequality is obvious, there is equality here.
$\square$

The validity of inequality (\ref{sub-add}) for arbitrary
$\omega\in\mathfrak{S}(\mathcal{H}\otimes\mathcal{K})$ seems to be
substantially stronger than the equivalent properties in theorem
2. This property is called \textit{subadditivity} of the
$\chi$-function for the channels $\Phi$ and $\Psi$. By using
arguments from the proof of theorem 2 it is easy to see that
subadditivity of the $\chi$-function for the channels $\Phi$ and
$\Psi$ is equivalent to validity of equality (\ref{addit}) for
\textit{arbitrary} subsets
$\mathcal{A}\subseteq\mathfrak{S}(\mathcal{H})$ and
$\mathcal{B}\subseteq\mathfrak{S}(\mathcal{K})$.

By using proposition 6 below it is possible to show that
properties $(i)-(iii)$ in the above theorem are equivalent to
subadditivity of the $\chi$-function for the channels $\Phi$ and
$\Psi$ having the following property: $H(\Phi(\rho))<+\infty$ and
$H(\Psi(\sigma))<+\infty$ for arbitrary finite rank states
$\rho\in\mathfrak{S}(\mathcal{H})$ and
$\sigma\in\mathfrak{S}(\mathcal{K})$.

We see later (proposition 7) that the set of quantum infinite
dimensional channels for which the subadditivity of the
$\chi$-function holds is nontrivial.

\textit{Remark 4.} By theorem 1 in \cite{H-Sh} the subadditivity
of the $\chi$-function for arbitrary finite dimensional channels
$\Phi$ and $\Psi$ is equivalent to validity of inequality
(\ref{super-add}) for arbitrary state
$\omega\in\mathfrak{S}(\mathcal{H}\otimes\mathcal{K})$, which
implies additivity of the minimal output entropy
\begin{equation}\label{add-min-entr}
\inf\limits_{\omega\in\mathfrak{S}(\mathcal{H}\otimes\mathcal{K})}H(\Phi\otimes\Psi(\omega))=
\inf\limits_{\rho\in\mathfrak{S}(\mathcal{H})}H(\Phi(\rho))
+\inf\limits_{\sigma\in\mathfrak{S}(\mathcal{K})}H(\Psi(\sigma))
\end{equation}
for these channels. This follows from the inequality
\begin{equation}\label{min-entr}
\begin{array}{c}
 H(\Phi\otimes\Psi(\omega))\geq\hat{H}_{\Phi\otimes\Psi}(\omega)\geq
\hat{H}_{\Phi} (\omega^{\mathcal{H}} )+\hat{H}_{\Psi}
(\omega^{\mathcal{K}})\\\\\geq\inf\limits_{\rho\in\mathfrak{S}(\mathcal{H})}H(\Phi(\rho))
+\inf\limits_{\sigma\in\mathfrak{S}(\mathcal{K})}H(\Psi(\sigma))
\end{array}
\end{equation}
valid for arbitrary state
$\omega\in\mathfrak{S}(\mathcal{H}\otimes\mathcal{K})$ for which
inequality (\ref{super-add}) holds.

In contrast to this in the infinite dimensional case we can not
prove the above implication (without some additional assumptions).
The problem consists in existence of pure states in
$\mathfrak{S}(\mathcal{H}\otimes\mathcal{K})$ with infinite
entropies of partial traces, which can be called
\textit{superentangled}. To show this note first that the
monotonicity property of the relative entropy \cite{L-2} provides
the following inequality
$$
\begin{array}{c}
H(\omega^{\mathcal{H}})+H(\omega^{\mathcal{K}})-H(\omega)\\\\=
H\left(\omega\,\|\,\omega^{\mathcal{H}}\otimes\omega^{\mathcal{K}}\right)
\geq
H\left(\Phi\otimes\Psi(\omega)\|\,\Phi\left(\omega^{\mathcal{H}}\right)\otimes\Psi\left(\omega^{\mathcal{K}}\right)\right)
\\\\= H\left(\Phi\left(\omega^{\mathcal{H}}\right)\right)
+H\left(\Psi\left(\omega^{\mathcal{K}}\right)\right)-H(\Phi\otimes\Psi(\omega)),
\end{array}
$$
which shows that
$H(\omega^{\mathcal{H}})=H(\omega^{\mathcal{K}})<+\infty$ implies
$H(\Phi(\omega^{\mathcal{H}}))<+\infty$ and
$H(\Psi(\omega^{\mathcal{K}}))<+\infty$ for arbitrary pure state
$\omega\in\mathfrak{S}(\mathcal{H}\otimes\mathcal{K})$ with finite
output entropy $H(\Phi\otimes\Psi(\omega))$. By this and theorem 2
the subadditivity of the $\chi$-function for arbitrary infinite
dimensional channels $\Phi$ and $\Psi$ implies validity of
inequality (\ref{super-add}) and hence validity of inequality
(\ref{min-entr}) for all pure states $\omega$ such that
$H(\omega^{\mathcal{H}})=H(\omega^{\mathcal{K}})<+\infty$ and
$H(\Phi\otimes\Psi(\omega))<+\infty$. So, if we considered only
such states $\omega$ in the calculation of the minimal output
entropy for the channel $\Phi\otimes\Psi$ we would obtain that it
is equal to the sum of
$\inf_{\rho\in\mathfrak{S}(\mathcal{H})}H(\Phi(\rho))$ and
$\inf_{\sigma\in\mathfrak{S}(\mathcal{K})}H(\Psi(\sigma))$, but
this additivity can be (probably) broken by taking into account
superentangled states.

\section{Generalization of the additivity conjecture}

The main aim of this section is to show that the conjecture of
additivity of the $\chi$-capacity for arbitrary finite dimensional
channels implies the additivity of the $\chi$-capacity for
arbitrary infinite dimensional channels with arbitrary
constraints.

It is convenient to introduce the following notation. The channel
$\Phi$ is
\begin{itemize}
  \item FF-channel if $\dim\mathcal{H}<+\infty$ and
$\dim\mathcal{H}'<+\infty$;
  \item FI-channel if $\dim\mathcal{H}<+\infty$ and
$\dim\mathcal{H}'\leq+\infty$.
 \end{itemize}

Speaking about quantum channel $\Phi$ without reference to FF or
FI we will assume that $\dim\mathcal{H}\leq+\infty$ and
$\dim\mathcal{H}'\leq+\infty$.

Let
$\Phi:\mathfrak{S}(\mathcal{H})\mapsto\mathfrak{S}(\mathcal{H}')$
be an arbitrary channel such that $\dim\mathcal{H}'=+\infty$ and
$P'_{n}$ be a sequence of finite rank projectors in $\mathcal{H}'$
increasing to $I_{\mathcal{H}'}$ and
$\mathcal{H}'_{n}=P'_{n}(\mathcal{H}')$. Consider the channel
\begin{equation}\label{Phi-n}
\Phi_{n}(\rho)=P'_{n}\Phi(\rho)P'_{n}+
\left(\mathrm{Tr}(I_{\mathcal{H}'}-P'_{n})\Phi(\rho)\right)\tau_{n}
\end{equation}
from $\mathfrak{S}(\mathcal{H})$ into
$\mathfrak{S}(\mathcal{H}'_{n}\oplus\mathcal{H}''_{n})\subset\mathfrak{S}(\mathcal{H}')$,
where $\tau_{n}$ is a pure state in some finite dimensional
subspace $\mathcal{H}''_{n}$ of
$\mathcal{H}'\ominus\mathcal{H}'_{n}$. If
$\dim\mathcal{H}'<+\infty$ we will assume that $\Phi_{n}=\Phi$ for
all $n$. Note that for arbitrary FI-channel $\Phi$ the
corresponding channel $\Phi_{n}$ is a FF-channel for all $n$.

For arbitrary channel
$\Psi:\mathfrak{S}(\mathcal{K})\mapsto\mathfrak{S}(\mathcal{K}')$
we will consider the sequences $\Phi_{n}$ and
$\Phi_{n}\otimes\Psi$ of channels as approximations for the
channels $\Phi$ and $\Phi\otimes\Psi$ correspondingly. Despite the
discontinuity of the $\chi$-capacity as a function of a channel in
the infinite dimensional case the following result is valid.

\textbf{Lemma 5.} \textit{Let $\Phi$ and $\Psi$ be arbitrary
channels. If subadditivity of the $\chi$-function holds for the
channel $\Phi_{n}$ defined by (\ref{Phi-n}) and the channel $\Psi$
for all $n$ then subadditivity of the $\chi$-function holds for
the channels $\Phi$ and $\Psi$.}

\textit{Proof.} The channel $\Phi_{n}$ can be represented as the
composition $\Pi_{n}\circ\Phi$ of the channel $\Phi$ with the
channel
$\Pi_{n}:\mathfrak{S}(\mathcal{H}')\mapsto\mathfrak{S}(\mathcal{H}_{n}'\oplus\mathcal{H}_{n}'')$
defined by
$$
\Pi_{n}(\rho')=P'_{n}\rho'P'_{n}+
\left(\mathrm{Tr}(I_{\mathcal{H}'}-P'_{n})\rho'\right)\tau_{n}.
$$

Proposition 4 implies
$$
\chi_{\Phi_{n}}(\rho)=\chi_{\Pi_{n}\circ\Phi}(\rho)\leq
\chi_{\Phi}(\rho),\quad\forall\rho\in
\mathfrak{S}(\mathcal{H}),\quad\forall n\in\mathbb{N}.
$$
Since
$$
\lim_{n\mapsto+\infty}\Phi_{n}(\rho)=\Phi(\rho),\quad\forall\rho\in
\mathfrak{S}(\mathcal{H})
$$
it follows from theorem 1 that
$$
\liminf_{n\mapsto+\infty}\chi_{\Phi_{n}}(\rho)\geq\chi_{\Phi}(\rho),\quad\forall\rho\in
\mathfrak{S}(\mathcal{H}).
$$
The two above inequalities imply
\begin{equation}\label{chi-n-1}
\lim_{n\mapsto+\infty}\chi_{\Phi_{n}}(\rho)=\chi_{\Phi}(\rho),\quad\forall\rho\in
\mathfrak{S}(\mathcal{H}).
\end{equation}

It is easy to see that
$$
\begin{array}{c}
\Phi_{n}\otimes\Psi(\omega)=(P'_{n}\otimes
I_{\mathcal{K}'})\cdot(\Phi\otimes\Psi(\omega))\cdot(P'_{n}\otimes
I_{\mathcal{K}'})\\\\+\,
\tau_{n}\otimes\mathrm{Tr}_{\mathcal{H}'}\left(((I_{\mathcal{H}'}-P'_{n})\otimes
I_{\mathcal{K}'})\cdot(\Phi\otimes\Psi(\omega))\right),
\quad\forall\omega\in\mathfrak{S}(\mathcal{H}\otimes\mathcal{K}).
\end{array}
$$
Hence
$$
\lim_{n\mapsto+\infty}\Phi_{n}\otimes\Psi(\omega)=\Phi\otimes\Psi(\omega),
\quad\forall\omega\in\mathfrak{S}(\mathcal{H}\otimes\mathcal{K}),
$$
and by theorem 1 we have
\begin{equation}\label{chi-n-2}
\liminf_{n\mapsto+\infty}\chi_{\Phi_{n}\otimes\Psi}(\omega)\geq\chi_{\Phi\otimes\Psi}(\omega),
\quad\forall\omega\in \mathfrak{S}(\mathcal{H}\otimes\mathcal{K}).
\end{equation}

By the assumption
\[
\chi_{\Phi_{n}\otimes\Psi}(\omega)\leq
\chi_{\Phi_{n}}(\omega^{\mathcal{H}})+\chi_{\Psi}(\omega^{\mathcal{K}}),
\quad\forall\omega\in
\mathfrak{S}(\mathcal{H}\otimes\mathcal{K}),\quad\forall
n\in\mathbb{N}.
\]
This, (\ref{chi-n-1}) and (\ref{chi-n-2}) imply
\[
\chi_{\Phi\otimes\Psi}(\omega)\leq
\chi_{\Phi}(\omega^{\mathcal{H}})+\chi_{\Psi}(\omega^{\mathcal{K}}),
\quad\forall\omega\in \mathfrak{S}(\mathcal{H}\otimes\mathcal{K}).
\,\square
\]

\textbf{Proposition 5.} \textit{Subadditivity of the
$\chi$-function for all FF-channels implies subadditivity of the
$\chi$-function for all FI-channels.}

\textit{Proof.} This can be proved by double application of lemma
5. First, we prove the subadditivity  of the $\chi$-function for
any two channels, when one of them is of FI-type while another is
of FF-type. Second, we remove FF restriction from the last
channel. $\square$

Now we will turn to channels with infinite dimensional input
quantum system. We will use the following notion of subchannel.

\textbf{Definition 3.} \textit{The restriction of a channel $\Phi
:\mathfrak{S}(\mathcal{H})\mapsto \mathfrak{S}(\mathcal{H}^{\prime
})$ to the set of states with support contained in a subspace
$\mathcal{H}_{0}$ of the space $\mathcal{H}$ is called subchannel
$\Phi_{0}$ of the channel $\Phi$, corresponding to the subspace
$\mathcal{H}_{0}$.}

It is easy to see that subadditivity of the $\chi$-function for
the channels $\Phi$ and $\Psi$ implies subadditivity of the
$\chi$-function for arbitrary subchannels $\Phi_{0}$ and
$\Psi_{0}$ of the channels $\Phi$ and $\Psi$. The properties of
the $\chi$-function established in section 4 make possible to
prove the following important result.

\textbf{Proposition 6.} \textit{Let
$\Phi:\mathfrak{S}(\mathcal{H})\mapsto\mathfrak{S}(\mathcal{H}')$
and
$\Psi:\mathfrak{S}(\mathcal{K})\mapsto\mathfrak{S}(\mathcal{K}')$
be arbitrary channels. Subadditivity of the $\chi$-function for
any two FI-subchannels of the channels $\Phi$ and $\Psi$ implies
subadditivity of the $\chi$-function for the channels $\Phi$ and
$\Psi$.}

\textit{Proof.} It is sufficient to consider the case
$\dim\mathcal{H}=+\infty$, $\dim\mathcal{K}\leq+\infty$. Let
$\omega$ be an arbitrary state in
$\mathfrak{S}(\mathcal{H}\otimes\mathcal{K})$. Let
$\{|\varphi_{k}\rangle\}_{k=1}^{+\infty}$ and
$\{|\psi_{k}\rangle\}_{k=1}^{\dim\mathcal{K}}$ be ONB of
eigenvectors of the compact positive operators
$\omega^{\mathcal{H}}$ and $\omega^{\mathcal{K}}$ such that the
corresponding sequences of eigenvalues are nonincreasing. Let
$P_{n}=\sum_{k=1}^{n}|\varphi_{k}\rangle\langle\varphi_{k}|$ and
$Q_{n}=\sum_{k=1}^{n}|\psi_{k}\rangle\langle\psi_{k}|$. In the
case $\dim\mathcal{K}<+\infty$ we will assume
$Q_{n}=I_{\mathcal{K}}$ for all $n\geq\dim\mathcal{K}$. The
nondecreasing sequences $\{P_{n}\}$ and $\{Q_{n}\}$ of finite rank
projectors converge to $I_{\mathcal{H}}$ and to $I_{\mathcal{K}}$
correspondingly in the strong operator topology. Let
$\mathcal{H}_{n}=P_{n}(\mathcal{H})$ and
$\mathcal{K}_{n}=Q_{n}(\mathcal{K})$.

Consider the sequence of states
$$\omega_{n}=
(\mathrm{Tr}\left((P_{n}\otimes
Q_{n})\cdot\omega\right))^{-1}(P_{n}\otimes Q_{n})\cdot\omega\cdot
(P_{n}\otimes Q_{n}),
$$
which are well defined for all $n$ by the choice  of the
projectors $P_{n}$ and $Q_{n}$. Since obviously
\begin{equation}\label{chi-nn-1}
\lim_{n\rightarrow+\infty}\omega_{n}=\omega
\end{equation}
proposition 3 implies
\begin{equation}\label{chi-nn-2}
\liminf_{n\rightarrow+\infty}\chi_{\Phi\otimes\Psi}(\omega_{n})
\geq\chi_{\Phi\otimes\Psi}(\omega).
\end{equation}

The next part of the proof is based on the following operator
inequalities
\begin{equation}\label{key-ineq}
\lambda_{n}\omega_{n}^{\mathcal{H}}\leq\omega^{\mathcal{H}},\quad
\lambda_{n}\omega_{n}^{\mathcal{K}}\leq\omega^{\mathcal{K}},\quad
\mathrm{where}\quad \lambda_{n}=\mathrm{Tr}\left((P_{n}\otimes
Q_{n})\cdot\omega\right).
\end{equation}

Let us prove the first inequality. By the choice of $P_{n}$ and
due to $\mathrm{supp}\omega_{n}^{\mathcal{H}}\subseteq
\mathcal{H}_{n}$ it is sufficient to show that
$\lambda_{n}\omega_{n}^{\mathcal{H}}\leq
P_{n}\omega^{\mathcal{H}}$. Let $\varphi\in \mathcal{H}_{n}$. By
definition of partial trace
$$
\begin{array}{c}
 \langle\varphi|\lambda_{n}\omega_{n}^{\mathcal{H}}|\varphi\rangle=
\sum\limits_{k=1}^{\dim\mathcal{K}}\langle\varphi\otimes\psi_{k}|P_{n}\otimes
Q_{n}\cdot\omega\cdot P_{n}\otimes
Q_{n}|\varphi\otimes\psi_{k}\rangle\\\\=
\sum\limits_{k=1}^{m}\langle\varphi\otimes\psi_{k}|\omega|\varphi\otimes\psi_{k}\rangle\leq
\sum\limits_{k=1}^{\dim\mathcal{K}}\langle\varphi\otimes\psi_{k}|\omega|\varphi\otimes\psi_{k}\rangle=
\langle\varphi|\omega^{\mathcal{H}}|\varphi\rangle,
\end{array}
$$
where $m=\min\{n,\dim\mathcal{K}\}$. The second inequality is
proved by the same way.

By using (\ref{chi-nn-1}) and applying corollary 4 due to
(\ref{key-ineq}) we obtain
\begin{equation}\label{chi-nn-3}
\lim_{n\rightarrow+\infty}\chi_{\Phi}(\omega_{n}^{\mathcal{H}})
=\chi_{\Phi}(\omega^{\mathcal{H}})\quad \mathrm{and}\quad
\lim_{n\rightarrow+\infty}\chi_{\Psi}(\omega_{n}^{\mathcal{K}})
=\chi_{\Psi}(\omega^{\mathcal{K}}).
\end{equation}

For each $n$ the $\{\omega^{\mathcal{H}}_{n}\}$-constrained
channel $\Phi$ and the $\{\omega^{\mathcal{K}}_{n}\}$-constrained
channel $\Psi$ can be considered as FI-subchannels of the channels
$\Phi$ and $\Psi$ corresponding to the subspaces $\mathcal{H}_{n}$
and $\mathcal{K}_{n}$. Hence by the assumption
$$
\chi_{\Phi\otimes\Psi}(\omega_{n})\leq
\chi_{\Phi}(\omega_{n}^{\mathcal{H}})+\chi_{\Psi}(\omega_{n}^{\mathcal{K}}),
\quad\forall n\in\mathbb{N}.
$$
This, (\ref{chi-nn-2}) and (\ref{chi-nn-3}) imply
\[
\chi_{\Phi\otimes\Psi}(\omega)\leq
\chi_{\Phi}(\omega^{\mathcal{H}})+\chi_{\Psi}(\omega^{\mathcal{K}}).
\,\square
\]

It is known, that additivity of the $\chi$-capacity for all
unconstrained FF-channels is equivalent to subadditivity of the
$\chi$-function for all FF-channels \cite{H-Sh},\cite{Sh-e-a-q}.
By combining this with proposition 5 and proposition 6 we obtain
the following extension of the additivity conjecture.

\textbf{Theorem 3.}  \textit{The additivity of the $\chi$-capacity
for all FF-channels implies additivity of the $\chi$-capacity for
all channels with arbitrary constraints.}

This theorem and theorem 2 implies the following result concerning
superadditivity of the convex closure of the output entropy for
infinite dimensional channels. Note that in the case of partial
trace channel the convex closure of the output entropy coincides
with the entanglement of formation (EoF).

\textbf{Corollary 5.} \textit{If inequality (\ref{super-add})
holds for all FF-channels $\Phi$ and $\Psi$ and all states
$\omega$ then inequality (\ref{super-add}) holds for all channels
$\Phi$ and $\Psi$ and all states $\omega$ such that
$H(\Phi(\omega^{\mathcal{H}}))<+\infty$ and
$H(\Psi(\omega^{\mathcal{K}}))<+\infty$.}

\textit{Proof.}  The validity of inequality (\ref{super-add}) for
two FF-channels $\Phi$ and $\Psi$ and for all states $\omega$ is
equivalent to subadditivity the $\chi$-function for these channels
\cite{H-Sh}. Hence the assumption of the corollary and theorem 3
imply subadditivity of the $\chi$-function for any channels,
which, by theorem 2, implies the validity of inequality
(\ref{super-add}) for all channels $\Phi$ and $\Psi$ and all
states $\omega$ such that $H(\Phi(\omega^{\mathcal{H}}))<+\infty$
and $H(\Psi(\omega^{\mathcal{K}}))<+\infty$.$\square$

\textit{Remark 5.} By combining Shor's theorem in \cite{Sh-e-a-q}
and theorem 3 we obtain that additivity of the minimal output
entropy (\ref{add-min-entr}) for all FF-channels implies
additivity of the $\chi$-capacity (\ref{addit}) for all channels
with arbitrary constraints. But due to existence of superentangled
states (see remark 4) we can not show that it implies additivity
of minimal output entropy for all channels. So, in the infinite
dimensional case the conjecture of additivity of the minimal
output entropy \textit{for all channels} seems to be substantially
stronger that the conjecture of additivity of the $\chi$-capacity
for all channels with arbitrary constraints.

Note that in contrast to proposition 5, proposition 6 relates the
subadditivity of the $\chi$-function for the initial channels with
the subadditivity of the $\chi$-function for its FI-subchannels
(not any FI-channels!). This makes it applicable for analysis of
individual channels as it is illustrated in the proof of
proposition 7 below.

We will use the following natural generalization of the notion of
entanglement breaking finite dimensional channel \cite{R}.

\textbf{Definition 4.} \textit{A channel
$\Phi:\mathfrak{S}(\mathcal{H})\mapsto\mathfrak{S}(\mathcal{H}')$
is called entanglement breaking if for an arbitrary Hilbert space
$\mathcal{K}$ and for an arbitrary state $\omega$ in
$\mathfrak{S}(\mathcal{H}\otimes\mathcal{K})$ the state
$\Phi\otimes \mathrm{Id}(\omega)$ lies in the closure of the
convex hull of all product states in
$\mathfrak{S}(\mathcal{H}'\otimes\mathcal{K})$, where
$\mathrm{Id}$ is the identity channel from
$\mathfrak{S}(\mathcal{K})$ onto itself.}

Generalizing the result in \cite{R} it is possible to show that a
channel
$\Phi:\mathfrak{S}(\mathcal{H})\mapsto\mathfrak{S}(\mathcal{H}')$
is entanglement-breaking if and only if it admits representation
$$
\Phi(\rho)=\int\limits_{X}\rho'(x)\mu_{\rho}(dx)
$$
where $X$ is a complete separable metric space, $\rho'(x)$ is a
Borel $\mathfrak{S}(\mathcal{H}')$-valued function on $X$ and
$\mu_{\rho}(A)=\mathrm{Tr}(\rho M(A))$ for any Borel $A\subset X$,
with $M$ positive operator valued measure on $X$ \cite{H-Sh-W}.

The following proposition is a generalization of proposition 2 in
\cite{H-Sh}.

\textbf{Proposition 7.} \textit{Let }$\Psi $\textit{\ be an
arbitrary channel. The subadditivity of the $\chi$-function holds
in each of the following cases: }

$\mathit{(i)}$ $\Phi $\textit{\ is a noiseless channel;}

$\mathit{(ii)}$ $\Phi $\textit{\ is an entanglement breaking
channel;}

$\mathit{(iii)}$ $\Phi $\textit{\ is a direct sum mixture
(cf.\cite{H-Sh}) of a noiseless channel and a channel $\Phi_{0}$
such that the subadditivity of the $\chi$-function holds for
$\Phi_{0}$ and $\Psi$ (in particular, an entanglement breaking
channel).}

\textit{Proof.} In the proof of each point of this proposition for
FF-channels  the finite dimensionality of the underlying Hilbert
spaces was used (cf.\cite{Sh-e-b-c},\cite{H-Sh}). The idea of this
proof consists in using our extension results (proposition 6 and
lemma 5).

$(i)$  Note that any FI-subchannel of an arbitrary noiseless
channel is a noiseless FF-channel. Hence by proposition 6 it is
sufficient to prove the subadditivity of the $\chi$-function for
arbitrary noiseless FF-channel $\Phi$ and arbitrary FI-channel
$\Psi$. But this can be done with the help of lemma 5. Indeed,
using this lemma with the noiseless FF-channel in the role of the
fixed channel $\Psi$ we can deduce the above assertion from the
subadditivity of the $\chi$-function for arbitrary two FF-channels
with one of them is a noiseless (proposition 2 in \cite{H-Sh}).

$(ii)$ Note that any FI-subchannel of an arbitrary entanglement
breaking channel is entanglement breaking. Hence by proposition 6
it is sufficient to prove the subadditivity of the $\chi$-function
for arbitrary entanglement breaking FI-channel $\Phi$ and
arbitrary FI-channel $\Psi$. Similar to the proof of $(i)$ this
can be done with the help of lemma 5, but in this case it is
necessary to apply this lemma twice. First we prove the
subadditivity of the $\chi$-function for arbitrary entanglement
breaking FI-channel $\Phi$ and arbitrary FF-channel $\Psi$ by
noting that any FF-channel $\Phi_{n}$, involved in lemma 5,
inherits the entanglement breaking property from the channel
$\Phi$ and using the subadditivity of the $\chi$-function for
arbitrary two FF-channels with one of them is an entanglement
breaking \cite{Sh-e-b-c}. Second, by using the result of the first
step we remove the FF restriction from another channel $\Psi$.

$(iii)$ Note that any FI-subchannel of the channel
$\Phi_{q}=q\mathrm{Id}\oplus (1-q)\Phi_{0}$ has the same structure
with FF-channel $\mathrm{Id}$ and FI-channel $\Phi_{0}$. By the
remark before proposition 6 subadditivity of the $\chi$-function
for the channels $\Phi_{0}$ and $\Psi$ implies subadditivity of
the $\chi$-function for arbitrary their subchannels. Hence by
proposition 6 it is sufficient to prove $(iii)$ for FI-channel
$\Phi_{q}$ and FI-channel $\Psi$.

Let  $\omega$ be a state in $\mathfrak{S}(\mathcal{H} \otimes
\mathcal{K})$ with  $\dim\mathcal{H}<+\infty$ and
$\dim\mathcal{K}<+\infty$. It follows that
$\chi_{\mathrm{Id}}(\omega^{\mathcal{H}})=H(\omega^{\mathcal{H}})<+\infty$.
By the established subadditivity of the $\chi$-function for
FF-channel $\mathrm{Id}$ and the FI-channel $\Psi$ and by the
assumed subadditivity of the $\chi$-function for FI-channel
$\Phi_{0}$ and the FI-channel $\Psi$ we have
$$
\chi _{\mathrm{Id}\otimes\Psi}(\omega )\leq \chi
_{\mathrm{Id}}(\omega^{\mathcal{H}})+\chi_{\Psi}(\omega^{\mathcal{K}})
\;\;\;\mathrm{and}\;\;\; \chi _{\Phi_{0}\otimes \Psi }(\omega
)\leq\chi_{\Phi_{0}}(\omega^{\mathcal{H}})+\chi_{\Psi}(\omega^{\mathcal{K}}).
$$
Using this and lemma 3 in \cite{H-Sh}\footnote{This lemma implies
that for arbitrary channels $\Phi_{1}$ and $\Phi_{2}$ from
$\mathfrak{S}(\mathcal{H})$ to $\mathfrak{S} (\mathcal{H}'_{1})$
and to $\mathfrak{S} (\mathcal{H}'_{2})$ correspondingly one has
\[
\chi _{q\Phi_{1}\oplus(1-q)\Phi_{2}}\left( \{\pi _{i},\rho
_{i}\}\right)=q\chi _{\Phi _{1}}\left( \{\pi _{i},\rho
_{i}\}\right)+(1-q)\chi _{\Phi _{2}}\left( \{\pi _{i},\rho
_{i}\}\right)
\]
for arbitrary ensemble $\{\pi _{i},\rho _{i}\}$ of states in
$\mathfrak{S}(\mathcal{H})$ and arbitrary $q\in[0;1]$.} we obtain
$$
\begin{array}{c}
\chi _{\Phi _q\otimes \Psi }(\omega)\leq q\chi
_{\mathrm{Id}\otimes \Psi }(\omega)+(1-q)\chi_{\Phi _0\otimes \Psi
}(\omega) \\
\\
\leq q\chi_{\mathrm{Id}}(\omega^{\mathcal{H}})+q\chi_{\Psi}
(\omega^{\mathcal{K}})+(1-q)\chi_{\Phi
_0}(\omega^{\mathcal{H}})+(1-q)\chi_{\Psi} (\omega^{\mathcal{K}})
\\
\\
=qH(\omega^{\mathcal{H}})+(1-q)\chi _{\Phi
_0}(\omega^{\mathcal{H}})+\chi_{\Psi}(\omega^{\mathcal{K}})=\chi
_{\Phi _q}(\omega^{\mathcal{H}})+\chi_{\Psi}
(\omega^{\mathcal{K}}),
\end{array}
$$
where the last equality follows from the existence of
approximating sequence of \textit{pure} state ensembles for the
$\{\omega^{\mathcal{H}}\}$-constrained FI-channel $\Phi_{0}$.
$\square$

\vspace{15pt}

\textbf{Acknowledgments.} The author is grateful to A. S. Holevo
for the idea of this work and permanent help and to M.B.Ruskai for
the useful remarks. The author also acknowledges support from QIS
Program, the Newton Institute, Cambridge, where this paper was
completed. The work was also partially supported by INTAS grant
00-738.


\begin{thebibliography}{99}

\bibitem{B&R} Bratteli O., Robinson D.W., "Operators algebras and quantum statistical mechanics";
Springer Verlag, New York-Heidelberg-Berlin, vol.I, 1979;

\bibitem{B&Ko} Bennett C.H., DiVincenzo D.P., Smolin J.A., Wootters W.K., "Mixed State Entanglement and Quantum Error
Correction", Phys. Rev. A 54, 3824-3851, 1996,  LANL e-print
quant-ph/9604024;

\bibitem{D-A}  Dell'Antonio G.F., "On the limits of sequences of normal states",
Commun. Pure Appl. Math. 20, 413-430, 1967;

\bibitem{Dev-1} Davies, E.B., "Information and Quantum Measurements",
              IEEE Trans.Inf.Theory 24, 596-599, 1978;

\bibitem{Dev-2} Davies, E.B., "Quantum theory of open systems",
              Academic Press, London, 1976;

\bibitem{Don} Donald M.J. "Further results on the relative entropy",
Math. Proc. Cam. Phil. Soc. 101, 363-373, 1987;

\bibitem{ESP} Eisert J., Simon C., Plenio M.B., "The quantification of entanglement in infinite-dimensional
quantum systems", J. Phys. A 35, 3911, 2002,  LANL e-print
quant-ph/0112064;

\bibitem{H-QI} Holevo, A.S., "Quantum coding theorems", Russian Math. Surveys, 53, N6,
 1295-1331, 1998, LANL e-print quant-ph/9809023;

\bibitem{H}  Holevo, A.S., "On quantum communication channels with
constrained inputs", LANL e-print quant-ph/9705054, 1997;

\bibitem{H-c-w-c} Holevo, A.S., "Classical capacities of quantum channels with
constrained inputs", Probability Theory and Applications, 48, N.2,
359-374, 2003, LANL e-print quant-ph/0211170;

\bibitem{H-Sh}  Holevo, A.S., Shirokov M.E., "On Shor's channel extension and
constrained channels", Commun. Math. Phys., 249, 417-430, 2004,
LANL e-print quant-ph/0306196, 2003;

\bibitem{H-Sh-2}  Holevo, A.S., Shirokov M.E., "Continuous ensembles and the $\chi$-capacity
of infinite dimensional channels", Probability Theory and
Applications, 50, N.1, 98-114, 2005, LANL e-print quant-ph/0408176;

\bibitem{H-W}  Holevo A.S., Werner R.F., "Evaluating capacities of Bosonic Gaussian channels",
Phys. Rev. A63, 032313; LANL e-print quant-ph/9912067, 1999;

\bibitem{H-Sh-W} Holevo A.S., Shirokov M.E., Werner R.F. "On the notion of entanglement in Hilbert
space", Russian Math. Surveys, 60, N.2, xxx, 2005, LANL e-print
quant-ph/0504204;

\bibitem{R} Horodecki M., Shor P.W., Ruskai, M.B. "General Entanglement Breaking Channels",
Rev. Math. Phys. 15, 629-641, 2003, LANL e-print quant-ph/0302031;

\bibitem{L-3}  Lindblad, G., "Entropy, Information and Quantum
Measurements", Comm. Math. Phys. 33, N.4, 305-322, 1973;

\bibitem{L}  Lindblad, G., "Expectation and Entropy Inequalities for Finite Quantum Systems",
Comm. Math. Phys. 39, N.2, 111-119, 1974;

\bibitem{L-2} Lindblad, G., "Completely Positive Maps and Entropy Inequalities",
Comm. Math. Phys. 40, N.2, 147-151, 1975;

\bibitem{O&P} Ohya M., Petz D., "Quantum Entropy and Its Use",
Texts and Monographs in Physics,  Berlin: Springer-Verlag, 1993;

\bibitem{P&B} Polovinkin E.S., Balashov M.V. "Elements of convex and strongly convex analysis",
2004 (In Russian);

\bibitem{S}  Sarymsakov, T.A., "Introduction to Quantum Probability
Theory", FAN, Tashkent, 1985, (In Russian);

\bibitem{Sch-West}  Schumacher, B., Westmoreland, M.D. "Sending Classical Information via Noisy Quantum Channels",
Phys. Rev. A 56, 131-138, 1997;

\bibitem{Sch-West-1}  Schumacher, B., Westmoreland, M.D. "Optimal signal
ensemble", Phys. Rev. A 63, 022308, 2001, LANL e-print
quant-ph/9912122;

\bibitem{Sh}  Shirokov M.E., "On the additivity conjecture for channels with
arbitrary constrains", LANL e-print quant-ph/0308168, 2003;

\bibitem{Sh-2}  Shirokov M.E., "On entropic quantities related to the classical capacity of infinite dimensional
quantum channels", LANL e-print quant-ph/0411091, 2004;

\bibitem{Sh-e-b-c}  Shor, P. W. "Additivity of the classical capacity of
entanglement breaking quantum channel", J.Math.Physics, 43,
4334-4340, 2002, LANL e-print quant-ph/0201149;

\bibitem{Sh-e-a-q}  Shor, P. W. "Equivalence of additivity questions in
quantum information theory", Comm. Math. Phys. 246, N.3, 453-472,
2004, LANL e-print quant-ph/0305035;

\bibitem{Simon}  Simon, B., "Convergence theorem for entropy", appendix in
Lieb E.H., Ruskai M.B., "Proof of the strong suadditivity of
quantum mechanical entropy", J.Math.Phys. 14, 1938, 1973;

\bibitem{W}  Wehrl, A., "General properties of entropy", Rev. Mod.
Phys. 50, 221-250, 1978.

\end{thebibliography}
\end{document}